\documentclass[12pt]{article}
\usepackage{fullpage, amsmath, amsthm, amsfonts, rotating, tabularx, array, float, booktabs, array, setspace, graphicx}
\usepackage{hyperref}
\usepackage{nameref}
\usepackage{multirow}
\usepackage{tikz}
\usepackage{tabularx}
\usetikzlibrary{shapes.geometric, arrows}
\usepackage{xcolor}

\usepackage[T1]{fontenc}
\usepackage{newtxtext}
\usepackage{newtxmath}

\usepackage{natbib}   

\newcommand{\bmx}{\mbox{\boldmath $x$}}

\definecolor{BlueGray}{RGB}{102, 153, 204}  

\title{Automatic Variance Adjustment for Small Area Estimation}
\author{Jon Wakefield$^{1,2}$, Jitong Jiang$^3$, Yunhan Wu$^4$\\ \\
$^1$Department of Statistics, University of Washington, Seattle, USA\\ $^2$Department of Biostatistics, University of Washington, Seattle, USA\\
$^3$Department of Biostatistics and Bioinformatics, Emory University, Atlanta, USA\\
$^4$Oak Ridge National Laboratory, Oak Ridge, USA}

\date{}                                           

\begin{document}
\newcommand{\shbox}[1]{\begin{center} \shadowbox{#1} \end{center}}

\def \N{\mbox{N}}
\def \v{\mbox{var}}
\def \E{\mbox{E}}
\def \c{\mbox{cov}}

\newcommand{\bA}{\mbox{\boldmath $A$}}
\newcommand{\bB}{\mbox{\boldmath $B$}}
\newcommand{\bC}{\mbox{\boldmath $C$}}
\newcommand{\bD}{\mbox{\boldmath $D$}}
\newcommand{\bE}{\mbox{\boldmath $E$}}
\newcommand{\bF}{\mbox{\boldmath $F$}}
\newcommand{\bG}{\mbox{\boldmath $G$}}
\newcommand{\bH}{\mbox{\boldmath $H$}}
\newcommand{\bI}{\mbox{\boldmath $I$}}
\newcommand{\bJ}{\mbox{\boldmath $J$}}
\newcommand{\bK}{\mbox{\boldmath $K$}}
\newcommand{\bL}{\mbox{\boldmath $L$}}
\newcommand{\bM}{\mbox{\boldmath $M$}}
\newcommand{\bN}{\mbox{\boldmath $N$}}
\newcommand{\bO}{\mbox{\boldmath $O$}}
\newcommand{\bP}{\mbox{\boldmath $P$}}
\newcommand{\bQ}{\mbox{\boldmath $Q$}}
\newcommand{\bR}{\mbox{\boldmath $R$}}
\newcommand{\bS}{\mbox{\boldmath $S$}}
\newcommand{\bT}{\mbox{\boldmath $T$}}
\newcommand{\bU}{\mbox{\boldmath $U$}}
\newcommand{\bV}{\mbox{\boldmath $V$}}
\newcommand{\bW}{\mbox{\boldmath $W$}}
\newcommand{\bX}{\mbox{\boldmath $X$}}
\newcommand{\bY}{\mbox{\boldmath $Y$}}
\newcommand{\bZ}{\mbox{\boldmath $Z$}}

\newcommand{\ba}{\mbox{\boldmath $a$}}
\newcommand{\bb}{\mbox{\boldmath $b$}}
\newcommand{\boldc}{\mbox{\boldmath $c$}}
\newcommand{\bd}{\mbox{\boldmath $d$}}
\newcommand{\bolde}{\mbox{\boldmath $e$}}
\newcommand{\boldf}{\mbox{\boldmath $f$}}
\newcommand{\bg}{\mbox{\boldmath $g$}}
\newcommand{\bh}{\mbox{\boldmath $h$}}
\newcommand{\boldi}{\mbox{\boldmath $i$}}
\newcommand{\bj}{\mbox{\boldmath $j$}}
\newcommand{\bk}{\mbox{\boldmath $k$}}
\newcommand{\bl}{\mbox{\boldmath $l$}}
\newcommand{\bm}{\mbox{\boldmath $m$}}
\newcommand{\bn}{\mbox{\boldmath $n$}}
\newcommand{\bo}{\mbox{\boldmath $o$}}
\newcommand{\bp}{\mbox{\boldmath $p$}}
\newcommand{\boldq}{\mbox{\boldmath $q$}}
\newcommand{\br}{\mbox{\boldmath $r$}}
\newcommand{\bolds}{\mbox{\boldmath $s$}}
\newcommand{\bt}{\mbox{\boldmath $t$}}
\newcommand{\bu}{\mbox{\boldmath $u$}}
\newcommand{\bv}{\mbox{\boldmath $v$}}
\newcommand{\bw}{\mbox{\boldmath $w$}}
\newcommand{\bx}{\mbox{\boldmath $x$}}
\newcommand{\by}{\mbox{\boldmath $y$}}
\newcommand{\bz}{\mbox{\boldmath $z$}}

\newcommand{\bmone}{\mbox{\bf 1}}
\newcommand{\bmzero}{\mbox{\bf 0}}

\newcommand{\balpha}{\mbox{\boldmath $\alpha$}}
\newcommand{\bbeta}{\mbox{\boldmath $\beta$}}
\newcommand{\blambda}{\mbox{\boldmath $\lambda$}}
\newcommand{\bgamma}{\mbox{\boldmath $\gamma$}}
\newcommand{\bdelta}{\mbox{\boldmath $\delta$}}
\newcommand{\bepsilon}{\mbox{\boldmath $\epsilon$}}
\newcommand{\bupsilon}{\mbox{\boldmath $\upsilon$}}
\newcommand{\btheta}{\mbox{\boldmath $\theta$}}
\newcommand{\bpsi}{\mbox{\boldmath $\psi$}}
\newcommand{\bphi}{\mbox{\boldmath $\phi$}}
\newcommand{\bnu}{\mbox{\boldmath $\nu$}}
\newcommand{\bmu}{\mbox{\boldmath $\mu$}}
\newcommand{\bOmega}{\mbox{\boldmath $\Omega$}}

\newcommand{\bomega}{\mbox{\boldmath $\bomega$}}
\newcommand{\bSigma}{\mbox{\boldmath $\Sigma$}}
\newcommand{\bmmSigma}{\mbox{\boldmath $\Sigma$}}

\newcommand{\bc}{\begin{center}}
\newcommand{\ec}{\end{center}}

\newcommand{\bi}{\begin{itemize}}
\newcommand{\ei}{\end{itemize}}

\newcommand{\be}{\begin{enumerate}}
\newcommand{\ee}{\end{enumerate}}

\newcommand{\bs}{\begin{slide*}}
\newcommand{\es}{\end{slide*}}

\newcommand{\bq}{ \begin{equation} }
\newcommand{\eq}{\end{equation}}

\maketitle

\begin{abstract}
Small area estimation (SAE) is a common endeavor and is used in a variety of disciplines. In low- and middle-income countries (LMICs), in which household surveys provide the most reliable and timely source of data, SAE is vital for highlighting disparities in health and demographic indicators. Weighted estimators are ideal for inference, but for fine geographical partitions in which there are insufficient data, SAE models are required. The most common approach is Fay-Herriot area-level modeling in which the data requirements are  a weighted estimate and an associated variance estimate. The latter can be undefined or unstable when data are sparse and so we propose a principled modification which is based on augmenting the available data with a prior sample from a hypothetical survey. This adjustment is generally available, respects the design and is simple to implement. We examine the empirical properties of the adjustment through simulation and illustrate its use with wasting data from a 2018 Zambian Demographic and Health Survey.
The modification is implemented as an automatic remedy in the {\tt R} package {\tt surveyPrev}, which provides a comprehensive suite of tools for conducing SAE in LMICs. 
\end{abstract}

\raggedright
\subsection*{Statement of Significance}

Fay-Herriot modeling is the most common approach to small area estimation and requires as input, as a minimum, a weighted estimate and an associated variance estimate. However, when sampling is sparse, relative to the geographical level at which estimates are required, its use can be hampered by instability or non-existence of the required variance estimates. We develop a procedure to produce a modified variance estimate that overcomes these difficulties and can be automatically applied within software. The latter is vital, since we are most concerned with situations in which the user does not have the time or expertise to carry out detailed variance modeling. Simulation experiments show the benefits of the approach. Our motivation is producing prevalence maps in low- and middle-income countries and we provide an analysis of wasting in children in Zambia in 2018, and illustrate that substantive conclusions can change when the variance modification is applied.

\subsection*{Data Availability Statement:} The Zambia 2018 DHS data can be accessed at \url{https://dhsprogram.com/}

\clearpage
\section{Introduction}

 Small area estimation (SAE), is defined as the task of ``producing reliable estimates of parameters of interest...for subpopulations (areas or domains) of a finite population for which samples of inadequate sizes or no samples are available'' \citep[p.~xxiii]{rao:molina:15}.
The generic problem is to produce estimates of finite population characteristics of interest over a set of areas. SAE is important in a variety of disciplines including global health, demography, education and economics, see \cite{rao:molina:15} for a range of examples.
In this paper, we focus on health and demographic indicators in low- and middle-income countries (LMICs). In LMICs, household surveys, such as the Demographic and Health Surveys (DHS), are the most reliable source of data.  Since 1984, more than 400 DHS surveys have been conducted in over 90 countries, creating a standardized and policy-relevant source of demographic and health information \citep{croft:18}. The DHS surveys use a stratified two-stage unequal probability cluster sampling design. This design is also used by the Multiple Indicator Cluster Sampling (MICS) program, which also carries out extensive surveys in LMICs \citep{khan2019multiple}.

A {\it direct estimator} for area $i$, for $i=1,\dots,I$, is one that only depends on response data from that area alone. A popular direct estimator is the weighted estimator that accounts for the sampling design by weighting responses using weights that acknowledge the sampling probabilities, and additionally may include adjustments for non-response and post-stratification. To accompany the estimator one may derive a design-based variance estimator. 
When data are sparse, some areas may produce a weighted estimate with unacceptably large uncertainty, while other areas may have no data at all. In these cases we may turn to SAE approaches that 
simultaneously model data from all areas in order to 
 increase precision, as compared to that of the direct estimates. The most commonly used approach is the two-stage Fay-Herriot model \citep{fay:herriot:79} that introduces area-specific random effects to link areas, with the possibility of including covariate information, to also aid in more precise predictions. 

In the original paper, the random effects were assumed to be independent and identically distributed (iid) normal.
Let $\widehat{\theta}^{\tiny{\text{w}}}_i$ represent a direct estimate of an area-level target parameter and 
$\widehat V_i$ be the sampling variance of $\widehat{\theta}_i$, which is estimated from data, using appropriate design-based variance formulas or resampling methods. 
The Fay-Herriot model is:
\begin{eqnarray}
\widehat{\theta}^{\tiny{\text{w}}}_i \mid \theta_i &\sim_{ind}& \mbox{N} ( \theta_i, \widehat V_{i}),\label{eq:fayherriot1}\\
 \theta_i&=&\alpha+ \bmx_i^{\mbox{\tiny{T}}}\boldsymbol{\beta} +  u_i , \label{eq:fayherriot2}\\
 u_i \mid \sigma^2_u&\sim_{iid} &\mbox{N}(0, \sigma^2_u),\quad i = 1, \ldots, I,\label{eq:fayherriot3}
\end{eqnarray}
where $\alpha$ is the intercept, $\bmx_i$ are area-level covariates with associated regression parameters $\boldsymbol{\beta}$, and $u_i $ represent between-area differences, which are modeled as random effects. We will describe the model from a Bayesian standpoint, since our implementation follows this path, and add a prior $\pi(\alpha,\boldsymbol{\beta},\sigma_u^2)$. The Fay-Herriot model implicitly acknowledges the sampling design through the use of sampling weights when computing the direct estimate and its standard error, $\widehat V^{1/2}_i$. 

In this paper we focus on estimating prevalences, since this is our motivation in LMICs (however, the methods we describe are applicable to other summaries also).
Often, the direct estimates may be transformed to make the normal approximation to the sampling distribution more accurate.  In particular, for each area $i$, we can define $\widehat{\theta}^{\,\tiny{\text{w}}}_i=h(\hat{p}_i^{\,\tiny{\text{w}}})$ where
$\hat{p}_i^{\tiny{\text{w}}}$ is the direct estimate of the prevalence and 
$h(\cdot)$ represents a transformation. The Fay-Herriot model can then be applied to the transformed $\hat{\theta}^{\,\tiny{\text{w}}}_i$ parameters with the resulting smoothed estimated being transformed back to the original scale. The sampling variance of the transformed $\hat{\theta}^{\,\tiny{\text{w}}}_i$ parameters can be approximated using the delta method. Transformations used with the Fay-Herriot model include the log, which was used in \cite{fay:herriot:79}, arcsin \citep{hirose_arc-sin_2023} and the logit \citep{mercer:etal:15}.

 The basic Fay-Herriot model assumes normally distributed iid area-level random effects, but the model may be easily extended to allow for random effects with other correlation structures. In particular, spatial and spatiotemporal covariance matrices may be used to smooth estimates across space and space-time, respectively. \cite{chung:datta:20} describe a range of spatial models including a conditionally autoregressive (CAR) model; they provide a comparison of the traditional Fay-Herriot model with spatial alternatives, finding that a spatial area-level model can improve estimation when good covariates are not available.  \cite{ghosh_generalized_1998} applied an intrinsic CAR (ICAR) prior \citep{besag:kooperberg:95} to the random effects, while other methods have focused on the use of simultaneous autoregressive (SAR) spatial models \citep{soton8165, petrucci2006smallcop, pratesi_small_2008, marhuenda_small_2013}.  The Besag-York-Mollié (BYM) model \citep{besag:york:mollie:91} that we have extensively used  consists of an unstructured iid normal random effect and a spatial ICAR random effect in each area. In our analyses in Section \ref{sec:zambia}, we adopt the reparameterization known as the BYM2 model \citep{riebler:etal:16}, in which the vector of random area effects $\boldsymbol{u}=[u_1,\dots,u_I]^{\mbox{\tiny{T}}}$ has structure,
\begin{equation}\label{eq:fayherriotBYM2}
\boldsymbol {u} = \sigma_u \left(\sqrt{1-\phi}\boldsymbol{e} + \sqrt{\phi} \boldsymbol{S}\right),
\end{equation}
where  $\sigma_u$ is the total standard deviation, $\phi$ is the proportion of the variance that is spatial, $\boldsymbol{e}=[e_1,\dots,e_I]^{\mbox{\tiny{T}}}$ is a vector of iid standard normal random variables and $\boldsymbol{S}=[S_1,\dots,S_I]^{\mbox{\tiny{T}}}$ follows a scaled ICAR prior, so that the geometric mean of the marginal variances of $S_i$ is equal to $1$, under a sum-to-zero constraint that is imposed to ensure identifiability when there is an intercept in the model \citep{rue:knorrheld:05}. This parameterization gives two interpretable parameters, $\sigma_u$ and $\phi$, for which hyperprior specification is relatively straightforward via penalized complexity (PC) priors \citep{simpson:etal:17}.

If there are just a small number of areas with no data then one may still fit Fay-Herriot models, treating these areas as having missing data. Spatial random effects models are particularly appealing in this regard, and the situation brightens considerably if there are strong associations with covariates. It is, however, very difficult to give guidelines on when the proportion of missing areas becomes too large to follow such a strategy.

The Fay-Herriot model is the most reliable approach when weighted estimates have unacceptably low precision, but the greater routine use of Fay-Herriot is hampered by unavailability of reliable variance estimates. 
In this paper, we propose a simple yet general approach to modifying variances of weighted estimates for subsequent use in Fay-Herriot models.

We are motivated by the task of providing SAE methods for LMICs, and work very closely with many National Statistics Offices (NSOs), UNICEF and the World Health Organization (WHO). 

For a recent review of SAE in a LMICs context, see \cite{wakefield25two}. 
Unstable sampling variances are a commonly encountered problem.
For example, in \cite{gardini2025mixture}, SAE models for DHS data in Bangladesh were considered, but design-based variances were unavailable/unstable in many areas. As an approximate solution, the design effect was approximated, and this was used as a method to estimate $\widehat V_i$. This method was not validated in any way, but the methods described in this paper would be applicable in this situation.

In general, to alleviate sampling variance difficulties, generalized variance functions \citep[Chapter 7]{wolter:07} may be used. Such approaches leverage the mean-variance relationship between $\theta^{\tiny{\text{w}}}_i$ and $\widehat{V}^{1/2}_i$, and/or incorporate  covariates \citep{otto:bell:95, mohadjer_hierarchical_2012, franco_applying_2013, liu_hierarchical_2014}. 
However, in LMICs, census covariates are less reliable, and so the use of covariates does not provide a general solution (though may be useful in particular scenarios, if the user has the time and resources to investigate covariate models). 
Uncertainty in the sampling variances may also be incorporated into the model by considering a joint model for the direct estimates and the associated sampling variance estimates \citep{you:chapman:06, maiti_prediction_2014, sugasawa_bayesian_2017,gao2023spatial}. As we have stressed, our aim is to develop an automated and fast method in LMICs and in addition to the problem of finding suitable covariates, the methods just highlighted are more difficult to implement and are not currently available in routine implementations.
We focus on the DHS design, but the methods we develop are universally applicable. 

The structure of this paper is as follows. Section \ref{sec:motivating} describes the 2018 DHS survey that was carried out in Zambia, and provides examples of where difficulties with variance estimation arise. In Section \ref{sec:methods} we describe variance estimation when data arise from surveys and our adjustment procedure, and then in Section \ref{sec:cluster} describe the specific modification that is relevant for the design used in DHS surveys. The procedure is demonstrated through simulation in Section \ref{sec:simulation} and in Section \ref{sec:zambia} we return to the Zambia example and show how the method works in practice. The paper concludes with a discussion in Section \ref{sec:discussion}. Additional simulation results and background derivations are relegated to the Appedices.

\section{Motivating Example: Zambia Demographic and Health Survey}\label{sec:motivating}

\begin{figure}
        \centering
        \includegraphics[width=15cm]{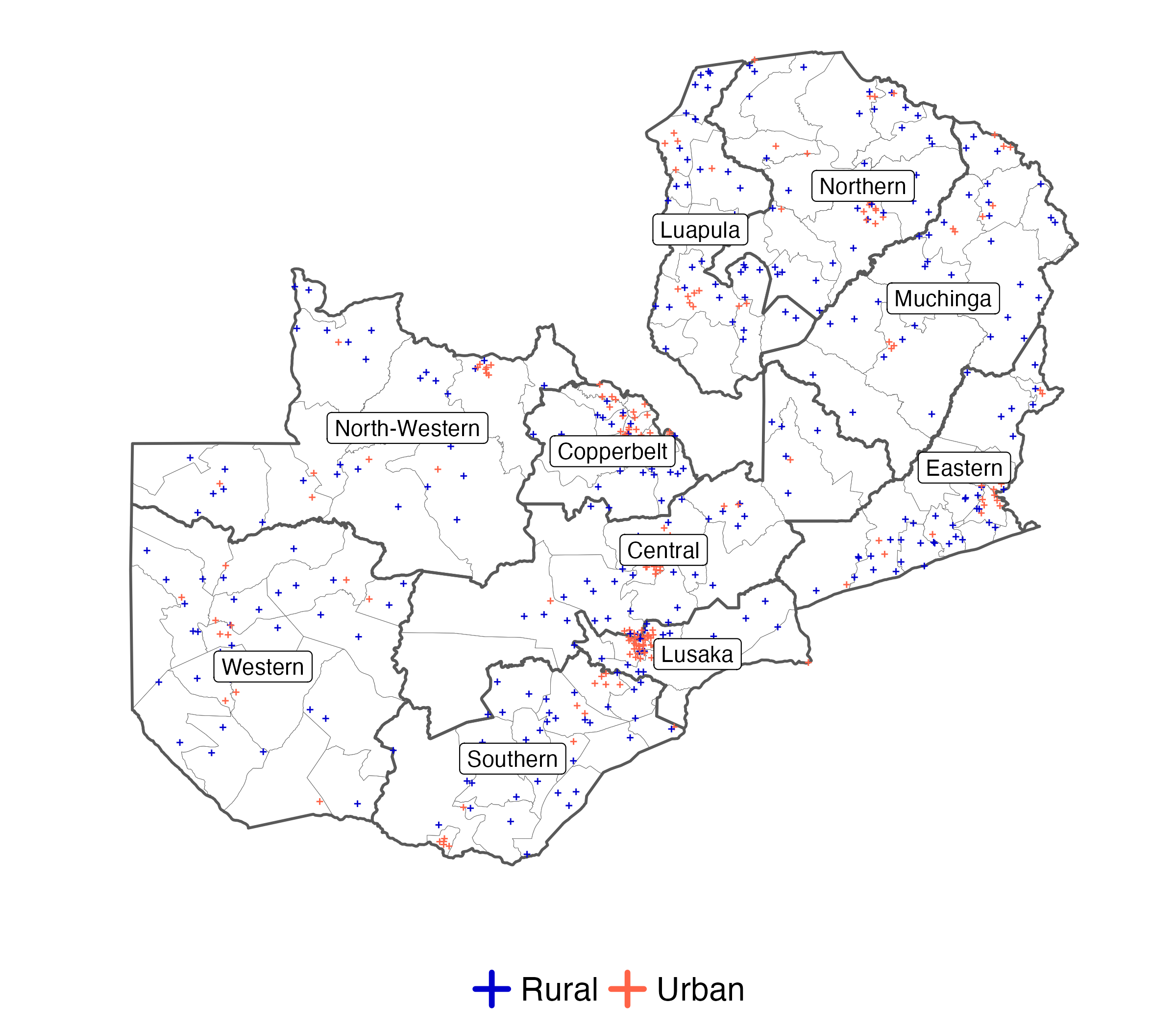}
        \caption{Locations of sampled urban and rural clusters in the 2018 Zambia DHS (jittered to preserve privacy) with Admin-1 and Admin-2 boundaries indicated, along with  Admin-1 labels.\label{fig:Zambia_map}}
\end{figure}

We begin with key definitions:~all countries are divided into principal administrative divisions, called Admin-1 regions, which are further subdivided into secondary administrative regions, called Admin-2 regions.  In our example, we wish to characterize variation in wasting in children across 10 provinces (Admin-1 areas) and 115 districts (Admin-2 areas) of Zambia, based on data from the 2018 DHS in which stratification is based on urban/rural crossed with Admin-1 areas.
The two stages of sampling are clusters, also called enumeration areas (EAs), within strata and households within clusters. For DHS, and more recent MICS surveys, responses are reported with their cluster location (subject to a random jitter that is added to the location for confidentiality reasons), so that all individuals in the cluster are reported to be located at the geographical location of the cluster (in practice, the reported locations are a jittered version of true locations, to aid in ensuring confidentiality). Figure \ref{fig:Zambia_map} maps the approximate locations of the 545 sampled clusters, indicating which were urban/rural in the original sampling frame based on the 2010 census. The sampling units are households, while the observation units are women. In our example, we consider wasting in children, and this information is obtained via the mothers.   All individuals within the same cluster receive the same design weight. Note that the sampling weights are normalized to obtain the final weight. The normalization process is done to obtain a total number of unweighted
cases equal to the total number of weighted cases at the national level for the total number of households,
women, and men. This weight is then multiplied by one million. Hence, totals cannot be estimated (without additional information) but ratios (such as the prevalence) are estimable.

Wasting (low weight-for-height) is a measure of acute under-nutrition and represents the failure to
receive adequate nutrition in the period immediately before the survey. The continuous measure is used to create a Z-score. Children whose Z-score is below
minus two standard deviations from the median of the reference
population (WHO Child Growth Standards) are considered thin (wasted), or acutely undernourished.  Wasting may result from
inadequate food intake or from a recent episode of illness or infection causing weight loss. 

Using the {\tt survey} package \citep{lumley:10} we calculate weighted (Hájek) estimates, along with their design-based variance estimates. 
The national
estimate of wasting for children under 5 years of age is 
0.042 (95\% interval: 0.036--0.048) with the urban prevalence of 0.049 (0.038--0.062) being 1.3 times greater than that in
rural areas, with the latter prevalence being 0.038 (0.032--0.045).

The Admin-1 and Admin-2 boundaries are shown in Figure \ref{fig:Zambia_map}, and we see a relatively large number of both urban and rural clusters in each Admin-1 area (as expected since these are planned domains), but sparser sampling in Admin-2 areas (unplanned domains).

\begin{figure} [!ht]
        \centering
        \includegraphics[clip, trim=0cm 8.5cm 0cm 8cm, width=1\linewidth]{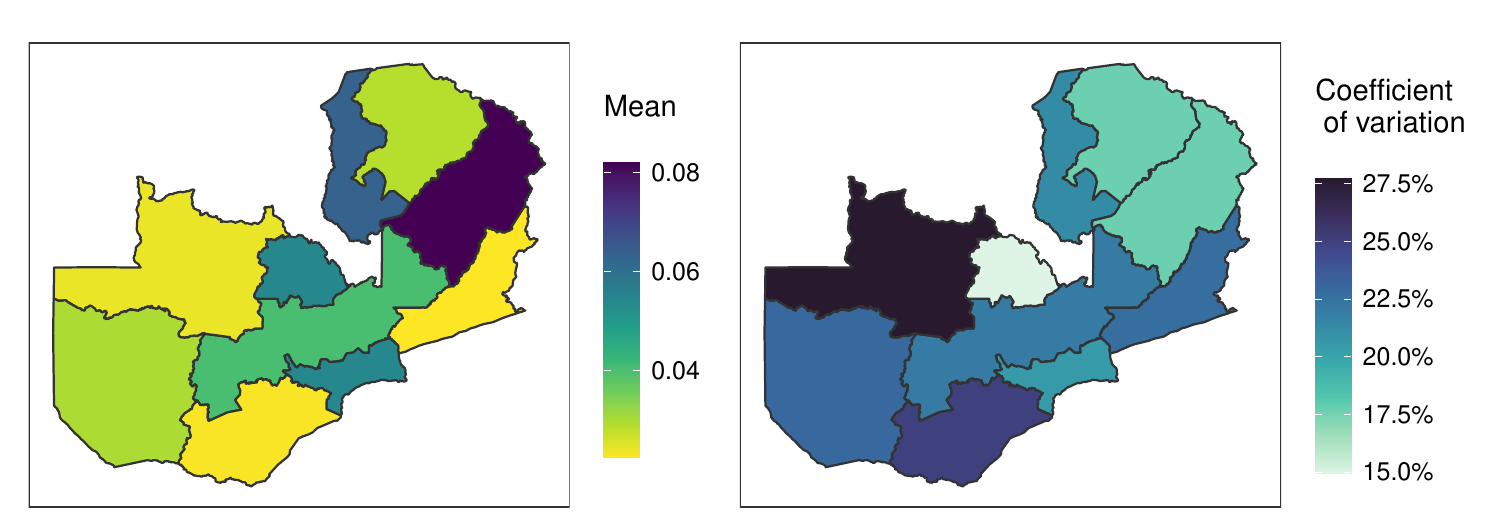}
        \caption{Admin-1 level wasting prevalence estimate (left) and coefficient of variation (right) in Zambia, based on DHS 2018 survey.}
        \label{fig:Zambia_wasting_direct_adm1}
\end{figure}

We turn now to domain estimation, again using weighted estimates.
At Admin-1, there are no issues, and Figure \ref{fig:Zambia_wasting_direct_adm1} gives a map of the prevalence (left) and the coefficient of variation (right).
Government agencies often require high precision when reporting domain estimates. For example, Statistics Canada, has guidelines \citep[Table 5]{cloutier2014aboriginal} for area-level estimates: for an area with a coefficient of variation (CV) of less than 16.7\% the estimate can be used without restriction, when the CV is above this but less than 33.3\%, it should be used with caution, and an estimate with a CV greater than 33.3\% is deemed too unreliable to be published. For the Zambia data, 27.7\% is the maximum CV of the (weighted) estimates over Admin-1 regions. 
However, only 9/115 of the Admin-2 areas have a CV
smaller than 16.7\% while 69/115 of the Admin-2 areas have a CV larger than 33.3\%. Consequently, SAE models are required to produce summaries that are reliable.

 To use the Fay-Herriot model, a variance estimate is required for each area. In Section \ref{sec:cluster} we give a closed form variance estimator that is appropriate for the DHS design. At the Admin-2 level, out of 115 areas, 3 areas have no clusters, and in 24 areas the variance is not estimable. In Section \ref{sec:cluster} we describe the exact details and data configurations that cause the variance formula to break. 
 
Hence, at the Admin-2 level, there is a need for  a modification of the variance estimates in order to use a Fay-Herriot model.  

\section{The General Augmentation Method}\label{sec:methods}

\subsection{Augmentation for Simple Random Sampling}

Our work is motivated by prevalence estimation in LMICs and so we focus upon the case of binary outcomes. It is straightforward to extend the method to continuous responses.
To motivate our variance adjustment we begin with a discussion of common approaches in the case of simple random sampling (SRS).
For SRS, suppose we have binary responses $y_k$, and we observe $y_{\cdot}=\sum_{k=1}^m y_k$ successes out of $m$ trials. The maximum likelihood estimator (MLE) is $\widehat{p}=y_{\cdot}/m$ with standard error $\sqrt{\widehat{p}(1-\widehat{p})/m}$. If $\widehat{p}=0$ or 1 the standard error is zero. The general problem of producing confidence intervals for the binomial $p$ has attracted many solutions and \cite{agresti:coull:98} provide an excellent summary in non-survey sampling situations. 

A Wald 100$\times (1-\alpha)$\% confidence interval of the form $\widehat p \pm z_{1-\alpha/2} \times \sqrt{\widehat{p}(1-\widehat{p})/m}$ (with $z_{1-\alpha/2}$ the normal quantile) performs poorly unless $m$ is quite large. The ``exact" solution based on inverting equal-tailed binomial tests of $H_0: p=p_0$ \citep{clopper1934use}, is known to be conservative. A score test, due to \cite{wilson:27}, is relatively accurate, and \cite{agresti:coull:98} show that its form can be approximated by the interval that results from a Bayesian approach with a Beta(2,2) prior.
An interesting perspective is provided by \cite{andersson2022approximate},  who examines the Wald interval in detail and delineates the reasons for its poor behavior for small $m$, including the skewness of the sampling distribution of $\widehat p$, and the dependence between the numerator and denominator in the Wald statistic. 

Our suggestion for adjusting problematic variances is based on extending a standard Bayesian approach to the complex survey design context.
In the SRS case, the binomial likelihood may be combined with a (conjugate) Beta$(a,b)$ prior to produce a Beta$(a+y_\cdot,b+m-y_\cdot)$ posterior from which a point estimate that moves  $\widehat{p}$ away from 0/1, and allows a posterior variance to be extracted and used in a modified Wald interval.  One way of picking $a$ and $b$ is to imagine a prior study with {\it phantom} data, that lead to
 $y_{\cdot}^{\tiny{\mbox{PH}}}
 $ 
 events of interest 
 from the $m^{\tiny{\mbox{PH}}}$ prior trials.
 In particular, we can take $a=y_\cdot^{\tiny{\mbox{PH}}}$ and $b=m^{\tiny{\mbox{PH}}}-y_{\cdot}^{\tiny{\mbox{PH}}}$ which gives prior mean of $\widehat p^{\tiny{\mbox{\,PH}}} = y_\cdot^{\tiny{\mbox{PH}}}/m^{\tiny{\mbox{PH}}}$. 

 As an alternative to direct specification of $a,b$ we can specify $\widehat p^{\tiny{\mbox{\,PH}}}$ and $m^{\tiny{\mbox{PH}}}$ and solve for $a,b$ via,
 \begin{eqnarray*}
 a&=&m^{\tiny{\mbox{PH}}} \widehat p^{\tiny{\mbox{\,PH}}} +1
 \qquad b=m^{\tiny{\mbox{\,PH}}} (1-\widehat p^{\tiny{\mbox{PH}}}) +1.
 \end{eqnarray*}
  The posterior mean is then the {\it augmented} estimator,
  $$\widehat p^{\tiny{\mbox{\,AUG}}} 
  = \frac{y_{\cdot}+y_{\cdot}^{\tiny{\mbox{PH}}}}{m+m_{\cdot}^{\tiny{\mbox{PH}}}}
  = \widehat{p} \times q_1 + \widehat p^{\tiny{\mbox{\,PH}}} \times q_2,$$
 where $\widehat p^{\tiny{\mbox{\,PH}}}= z^{\tiny{\mbox{PH}}}/m^{\tiny{\mbox{PH}}} $ is the prior mean, $q_1=m/(m+m^{\tiny{\mbox{PH}}})$ and $q_2=m^{\tiny{\mbox{PH}}}/(m+m^{\tiny{\mbox{PH}}})$.
  This approach is appealing since the estimator is consistent as $m \rightarrow \infty$ and the required inputs, a prior estimate and a prior sample size, are straightforward to think about. The question is how to adapt this approach to the survey setting, and we address this in the next section.
  
\subsection{Augmentation for Complex Sampling}

We would like to obtain variances that are well-defined, and interval estimates with close to nominal coverage, when the data arise from a complex survey.
In an extensive simulation study, \cite{franco2019comparative} compare a number of methods in the complex design setting, see also \cite{korn1998confidence} and \cite{dean2015evaluating}.
Many methods for improving coverage in the complex design setting use methods for SRS, but replace the sample size $m$ with a design-effect adjusted version. In out setting, this is less appealing, since to estimate the design effect the variance is required, and we are interested in situations in which the variance is not reliably estimated.

In this section we consider a generic survey sampling scenario in which binary observations $y_k$, for $k \in S$ are sampled with asscoated design weights $w_k=1/\pi_k$, where $\pi_k$ is the selection probability for unit $k$ of the sample. 
The Hájek estimator 
is,
$$\widehat p^{\tiny{\,\mbox{HJ}}} = \frac{\sum_{k \in S} w_{k} y_{k}}{ \sum_{k \in S} w_{k} }.$$
Consider the pseudo-likelihood \citep{binder:83,pfeffermann:etal:98},
$$
L^{\tiny{\mbox{PL}}}(p) = \prod_{k \in S}  \left[ ~p^{y_{k}}(1-p)^{1-y_{k}} \right]^{w_{k}}.
$$
and note that $\widehat p^{\tiny{\,\mbox{HJ}}}$ maximizes $L^{\tiny{\mbox{PL}}}(p)$.

Now define a {\it pseudo prior}, based on phantom data, $\{y_k, w_k, k \in S^{\tiny{\mbox{PH}}}\}$,
$$
\pi^{\tiny{\mbox{PH}}}(p) \propto \prod_{k \in S^{\tiny{\mbox{PH}}}}  \left[ ~p^{y_k}(1-p)^{1-y_k} \right]^{w_{k}} \times [p(1-p)]^{-1},
$$

which is a Beta$\left(\sum_{k \in S^{\tiny{\mbox{PH}}}} w_ky_k, \sum_{k \in S^{\tiny{\mbox{PH}}}} w_k(1-y_k) \right)$ distribution. 
 
The prior is, $$\widehat p^{\tiny{\mbox{\,PH}}} = \frac{\sum_{k \in S^{\tiny{\mbox{PH}}} } w_{k} y_{k}}{ \sum_{k \in S^{\tiny{\mbox{PH}}}} w_{k} }.$$ 
The  pseudo (augmented) posterior is,
$$
\pi^{\tiny{\mbox{AUG}}}(p)
\propto \prod_{k \in S^{\tiny{\mbox{AUG}}}}  \left[~ p^{y_{k}}(1-p)^{1-y_{k}} \right]^{w_{k}}
 \times [p(1-p)]^{-1}$$
where $S^{\tiny{\mbox{AUG}}} = S \cup S^{\tiny{\mbox{PH}}}$.
We refer to the expected value of the pseudo posterior as the (augmented) estimator:
$$ 
\widehat p^{\tiny{\mbox{\,AUG}}}  = \frac{\sum_{k \in S^{\tiny{\mbox{AUG}}}} w_{k} y_{k}}{\sum_{k \in S^{\tiny{\mbox{AUG}}}}  w_{k} }
=  \widehat p^{\tiny{\mbox{\,HJ}}} \times 
q_1  +  \widehat p^{\tiny{\mbox{\,PH}}} \times 
q_2.
$$
where
$$q_1 = \frac{\sum_{k \in S} w_{k}}{\sum_{k \in S} w_{k}+\sum_{k \in S^{\tiny{\mbox{PH}}}}  w_{k} },\qquad q_2 = \frac{\sum_{k \in S^{\tiny{\mbox{PH}}}} w_{k}}{\sum_{k \in S} w_{k}+\sum_{k \in S^{\tiny{\mbox{PH}}}}  w_{k} }.$$
Note that we just need to specify the prior mean, $\widehat p^{\tiny{\mbox{PH}}}$, and the sum of the weights in the prior sample, $\sum_{k \in S^{\tiny{\mbox{PH}}}}  w_{k}$. 
 We introduced the pseudo-prior as a mechanism to provide a regularization procedure. For evaluating the variance of $\widehat p^{\tiny{\,\mbox{HJ}}}$ we cannot simply use the 
posterior variance, since this does not account for the complex sampling. Instead, we need to use the relevant design-based variance formula \citep{binder:83} but now for the augmented data.
The posterior variance arising from the beta posterior is not the appropriate measure of uncertainty, since it does not appropriately account for the design.

To obtain the relevant variance for the augmented sample, we simply use the method that was used for the original data, but now with the augmented sample.
For many designs, a relevant closed-form variance formula is available, or more generally, one may use the jackknife.

\subsection{Augmentation in the Exponential Family}

In this section we illustrate that the above derivation holds for pseudo-likelihoods beyond the binomial.
Consider the one-parameter exponential family,

$$p(y \mid \theta )  = h(y) \exp\left[
s(y) \theta - M (\theta) 
\right]$$
where $h(y) \geq 0$, $s(y)$ is a real-valued function of the observation $y$ and 
$$M(\theta) = \log \left\{ \int
h(y) \exp[s(y)\theta] ~dy
\right\},$$
with $M'(\theta)=\mbox{E}[Y]=\mu(\theta)$.
Now suppose we collect data $y_k$,  under a sampling scheme with weights $w_k$, $k \in S$. Consider a linear exponential family, so that $s(y)=y$. The corresponding log pseudo-likelihood \citep{binder:83} is 
$$\log L^{\tiny{\mbox{PL}}}(\theta ) = \sum_{k \in S} w_k \log h(y_k) + \theta \sum_{k \in S} w_k y_k  - M(\theta) \sum_{k \in S} w_k  .$$
Maximizing yields the Hájek estimator,
$$M'(\widehat \theta) = \widehat \mu^{\tiny{\mbox{HJ}}} = \frac{\sum_{k \in S} w_k y_k}{\sum_{k \in S} w_k}.$$

To construct a prior we imagine phantom  observations, $\{y_k,w_k,k \in S^{\tiny{\mbox{PH}}}\}$, and use these to construct a weighted (log) conjugate prior:
$$\log \pi^{\tiny{\mbox{PH}}}( \theta ) = \sum_{k \in S^{\tiny{\mbox{PH}}} } w_k \log h(y_k) + \theta\sum_{k \in S^{\tiny{\mbox{PH}}}} w_k y_k  - M(\theta)\sum_{k \in S^{\tiny{\mbox{PH}}}} w_k  .$$
The log pseudo-posterior is,
$$\log \pi^{\tiny{\mbox{AUG}}}( \theta ) = \sum_{k \in S^{\tiny{\mbox{PH}}}\cup S} w_k \log h(y_k) + \theta \sum_{k \in S^{\tiny{\mbox{PH}}}\cup S} w_k y_k  - M(\theta) \sum_{k \in S^{\tiny{\mbox{PH}}}\cup S} w_k  .$$
The posterior mean is \citep{diaconis1979conjugate},
$$\widehat \mu^{\tiny{\mbox{AUG}}} = \frac{\sum_{k \in S^{\tiny{\mbox{PH}}} \cup S} w_k y_k}{\sum_{k \in S^{\tiny{\mbox{PH}}} \cup S} w_k } = \frac{\sum_{k \in S} w_k }{\sum_{k \in S^{\tiny{\mbox{PH}}}\cup S} w_k}\times\widehat  \mu ^{\tiny{\mbox{HJ}}} + \frac{\sum_{k \in S^{\tiny{\mbox{PH}}}} w_k }{\sum_{k \in S^{\tiny{\mbox{PH}}}\cup S} w_k}\times \widehat  \mu ^{\tiny{\mbox{PH}}}  
$$
 
Since the form of the estimator is identical to the Hájek estimator, with an augmented set of observations that includes the phantom prior observations, we can again use standard approaches to obtain the variance, for example, linearization or the jackknife.

\section{Adjustment for Stratified Two-Stage Unequal Probability Cluster Sampling}\label{sec:cluster}

\subsection{Weighted Estimator and its Variance}

In this section we describe the adjustment method for our motivating example, prevalence mapping in Zambia from the 2018 DHS, which uses stratified two-stage unequal probability cluster sampling. Recall that in the DHS (and most of the surveys carried out in LMICs) the stratification is usually Admin-1 areas crossed with urban/rural. For completeness we will consider planned (i.e.,~Admin-1) and unplanned (i.e.,~Admin-2) domains, even though the former will generally not have variance issues, unless we have a very rare outcome or this could occur for an indicator which the design did not have in mind, when stratum sizes were taken. Also, in extreme circumstances it may occur if there is substantial non-response, or clusters cannot be visited, perhaps for security reasons. For example, in the Nigeria 2018 DHS, states were strata and in Borno state, 11 of the 27 clusters were dropped due to high insecurity. 

We need to consider the urban/rural stratification of DHS, since the target areas generally contain both urban and rural segments. We let $h_1=1,\dots,H_1$ represent the Admin-1 areas that form one element of the stratification and $h_2=1,2$ represent the second urban/rural element. 
Let $y_{h_1h_2ck}=0/1$ be the binary response, and $w_{h_1h_2ck}$ be the associated design weight, for sampled unit $k$ in cluster $c$ of the stratum determined by $h_1$ crossed with $h_2$,  $c \in S_{h_1h_2}$,  $k \in S_{h_1h_2c}$, for $h_1=1,\dots,H_1$, $h_2=1,2$. 
Hence, $S_{h_1h_2}$ are the set of sampled clusters in the stratum defined by $h_1,h_2$ and $S_{h_1h_2c}$ are the set of individuals who are subsequently selected in cluster $c$.

We define notation that will allow us to give expressions for both planned and unplanned domains, indexed by $i$ with $U_i$ representing the population units in area $i$. For both planned and unplanned domains, area $i$ corresponds to exactly one Admin-1 area, $h_1[i]$, with $i=h_1[i]$ for planned domains.
Following standard practice \citep{sas2016sas}, define {\it extended domain variables},
\begin{eqnarray*}
    z_{ih_1h_2ck} &=& \left\{ \begin{array}{ll}
      I( k \in U_i ) y_{h_1[i]h_2ck}  &  \mbox{ i.e.,~if unit $k$ belongs to target domain $i$ 
      } 
      \\
      0   & \mbox{ otherwise. }
    \end{array}
    \right.\\
       v_{ih_1h_2ck} &=& \left\{ \begin{array}{ll}
      I( k \in U_i ) w_{h_1[i]h_2ck}  &  \mbox{ i.e.,~if unit $k$ belongs to target domain $i$ 
      } 
      \\
      0   & \mbox{ otherwise. }
    \end{array}
    \right. ,
\end{eqnarray*}
where $I(A)=1$ if  event $A$ occurs, and $=0$ otherwise.
For use later, define,
$$
v_{ih_1[i]h_2c\cdot} = \sum_{k \in S_{ih_1[i]h_2c}} v_{ih_1[i]h_2ck},
v_{ih_1[i]h_2\cdot \cdot} = \sum_{c \in S_{ih_1h_2}}v_{ih_1[i]h_2c\cdot},\qquad
v_{ih_1[i]\cdot\cdot\cdot}= \sum_{h_2=1}^2 v_{ih_1[i]h_2\cdot \cdot}.
$$
The Hájek estimator in strata $h_1,h_2$ can be derived as the pseudo MLE from the pseudo likelihood:
\begin{equation}\label{eq:PLcluster}
L^{\tiny{\mbox{PL}}}(p_{ih_1[i]h_2}) =  
\prod_{c \in S_{h_1[i]h_2} }
\prod_{k \in S_{h_1[i]h_2c}}  \left[ ~p_{ih_1[i]h_2}^{z_{ih_1[i]h_2ck}}(1-p_{ih_1[i]h_2})^{1-z_{ih_1[i]h_2ck}} \right]^{v_{ih_1[i]h_2ck}}.
\end{equation} 
Maximization yields,
\begin{equation}\label{eq:haaj}
\widehat p_{ih_1[i]h_2}^{\tiny{\mbox{\,HJ}}} 
= \frac{\sum_{c \in S_{h_1[i]h_2}} \sum_{k \in S_{h_1[i]h_2c}} v_{ih_1[i]h_2ck} z_{ih_1[i]h_2ck}}
{ \sum_{c \in S_{h_1[i]h_2}} \sum_{k \in S_{h_1[i]h_2c}} v_{ih_1[i]h_2ck}}
\end{equation}

The prevalence for area $i$ is,
\begin{equation}\label{eq:hajekh}
\widehat p_i^{\tiny{\mbox{\,HJ}}}=
\sum_{h_2=1}^2 q_{ih_1[i]h_2} \times \widehat p_{ih_1[i]h_2}^{\tiny{\mbox{\,HJ}}}
\end{equation}
where
$$
q_{ih_1[i]h_2} = 
\frac{
\sum_{c \in S_{h_1[i]h_2} }
\sum_{k \in S_{h_1[i]h_2c}} v_{ih_1[i]h_2ck}}
{\sum_{h_2=1}^2 \sum_{c \in S_{h_1[i]h_2} }
\sum_{k \in S_{h_1[i]h_2c}} 
v_{ih_1[i]h_2ck}}
= \frac{v_{ih_1[i]h_2 \cdot\cdot}}{v_{ih_{ih_1[i]\cdot\cdot\cdot}}}
.
$$

In the {\tt survey} package in {\tt R} the variance for stratified cluster sampling is calculated by taking a sequential view of multistage sampling in which at each stage an additional variance term is added \nocite{lumley:10} (Lumley, 2010, p.~41) -- in particular, equation (4.4.3) of \cite{sarndal:etal:92} is used for totals. This variance estimator does not have a closed form so  we use existing closed-form estimators, because we wish to obtain a form that can be extended to include phantom observations in a convenient fashion.

To define the variance of the estimator, for either planned (Admin-1) or unplanned (Admin-2) area $i$, we first define $ n_{h_1[i]h_2}$ as the number of clusters in strata $h_1[i],h_2$. Recall, if the areas are planned then area $i$ corresponds to Admin-1 strata $h_1[i]$, and if unplanned (Admin-2), then it is a subregion of $h_1[i]$. The variance of the estimator in (\ref{eq:hajekh}) is \citep[p.~9282--9283]{sas2016sas},
$$ \widehat{V}(\widehat p_i^{\tiny{\,\mbox{HJ}}}) =
\sum_{h_2=1}^2 \widehat{V}(\widehat p_{ih_1[i]h_2}^{\tiny{\,\mbox{HJ}}}),$$
where,
\begin{equation}\label{eq:sasvar}
 \widehat{V}(\widehat p_{ih_1[i]h_2}^{\tiny{\,\mbox{HJ}}}) = 
\frac{1}{v_{ih_1[i]\cdot\cdot\cdot}^2} \frac{n_{ih_1[i]h_2}}{n_{ih_1[i]h_2}-1}\sum_{c \in S_{ih_1[i]h_2}} \left[
v_{ih_1[i]h_2c\cdot} ( \widehat p_{ih_1[i]h_2c}^{\tiny{\,\mbox{HJ}}}-\widehat p_i^{\tiny{\,\mbox{HJ}}}) -\frac{1}{n_{ih_1[i]h_2}}
v_{ih_1[i]h_2\cdot\cdot}( \widehat p_{ih_1[i]h_2}^{\tiny{\,\mbox{HJ}}}-\widehat p_i^{\tiny{\,\mbox{HJ}}})
\right]^2 .
\end{equation}
and
$$ 
\widehat p_{ih_1[i]h_2}^{\tiny{\mbox{\,HJ}}} = \frac{\sum_{c \in S_{ih_1[i]h_2}} \sum_{k \in S_{ih_1[i]h_2c}} v_{ih_1[i]h_2ck} z_{ih_1[i]h_2ck}}{v_{ih_1[i]h_2\cdot\cdot}},\qquad \widehat p_{ih_1[i]h_2c}^{\tiny{\mbox{\,HJ}}} = \frac{\sum_{k \in S_{ih_1[i]h_2c}} v_{ih_1[i]h_2ck} z_{ih_1[i]h_2ck}}{v_{ih_1[i]h_2c\cdot}}.$$
In Appendix A, we give an explicit form that shows for the unplanned domain case, there is an extra term from clusters outside of the domain, that accounts for the additional variation due to the random number of observations that fall in the domain.

In Appendix B, we show that (\ref{eq:sasvar}) produces virtually identical variance estimates to the procedure used in the {\tt survey} package.

\subsection{The Augmented Estimator and its Variance}

To obtain the augmented variance estimator, first consider the pseudo-likelihood for the urban/rural strata specific prevalences within area $i$, $\{ p_{ih_1[i],h_2}, h_2=1,2\}$:
\begin{equation}\label{eq:PLcluster2}
L^{\tiny{\mbox{PL}}}( p_{ih_1[i]h_2=1}, p_{ih_1[i]h_2=2} ) = \prod_{h_2=1}^2
L_{h_2}^{\tiny{\mbox{PL}}}(p_{ih_1[i]h_2}),
\end{equation}
where  $L_{h_2}$ is given by (\ref{eq:PLcluster}).
We combine this pseudo-likelihood with the pseudo-prior 
$$\pi^{\tiny{\mbox{PH}}}( p_{ih_1[i],h_2=1},p_{ih_1[i],h_2=2} ) = \prod_{h_2=1}^2 \pi_{h_2}^{\tiny{\mbox{PH}}}(p_{ih_1[i]h_2})$$
where
\begin{equation}\label{eq:pp}
\pi_{h_2}^{\tiny{\mbox{PH}}}(p_{ih_1[i]h_2}) = 
\prod_{c \in S^{\tiny{\mbox{\,PH}}}_{ih_1[i]h_2}}
\prod_{k \in S^{\tiny{\mbox{\,PH}}}_{ih_1[i]h_2c}}
[ ~p_{ih_1[i]h_2}^{z_{ih_1[i]h_2ck}}(
1-p_{ih_1[i]h_2})^{1-z_{ih_1[i]h_2ck}}~]^{v_{ih_1[i]h_2ck}} [p_{ih_1[i]h_2}(1-p_{ih_1[i]h_2})]^{-1}.
\end{equation}
This prior is based on phantom clusters in each of the strata.
Combining (\ref{eq:PLcluster2}) and (\ref{eq:pp}) gives pseudo-posterior,
$$
\pi_{h_2}^{\tiny{\mbox{AUG}}}(p_{ih_1[i]h_2}) = 
\prod_{c \in S^{\tiny{\mbox{\,AUG}}}_{ih_1[i]h_2}}
\prod_{k \in S^{\tiny{\mbox{\,AUG}}}_{ih_1[i]h_2c}}
[ p_{ih_1[i]h_2}^{z_{ih_1[i]h_2ck}}(
1-p_{ih_1[i]h_2})^{1-z_{ih_1[i]h_2ck}}]^{v_{ih_1[i]h_2ck}}[p_{ih_1[i]h_2}(1-p_{ih_1[i]h_2})]^{-1},
$$
where $S^{\tiny{\mbox{\,AUG}}}_{ih_1[i]h_2} = S_{ih_1[i]h_2} \cup S^{\tiny{\mbox{\,PH}}}_{ih_1[i]h_2}$ and $S^{\tiny{\mbox{\,AUG}}}_{ih_1[i]h_2c} = S_{ih_1[i]h_2c} \cup S^{\tiny{\mbox{\,PH}}}_{ih_1[i]h_2c}$.
The posterior mean is,
$$
\widehat p_{ih_1[i]h_2}^{\tiny{\mbox{\,AUG}}}=
\frac{
\sum_{c \in S^{\tiny{\mbox{\,AUG}}}_{ih_1[i]h_2}} \sum_{k \in S^{\tiny{\mbox{\,AUG}}}_{ih_1[i]h_2c}
} v_{ih_1[i]h_2ck}z_{ih_1[i]h_2ck}}
{\sum_{c \in S^{\tiny{\mbox{\,AUG}}}_{ih_1[i]h_2}} \sum_{k \in S^{\tiny{\mbox{\,AUG}}}_{ih_1[i]h_2c}}v_{ih_1[i]h_2ck} },
$$
leading to augmented prevalence estimator (for planned or unplanned domains),
$$\widehat p_i^{\tiny{\mbox{\,AUG}}}=\sum_{h_2=1}^2 q_{ih_1[i]h_2} \times \widehat p_{ih_1[i]h_2}^{\tiny{\mbox{\,AUG}}}   = \frac{ \sum_{h_2=1}^2 \sum_{c \in S^{\tiny{\mbox{\,AUG}}}_{ih_1[i]h_2}} \sum_{k \in S^{\tiny{\mbox{\,AUG}}}_{ih_1[i]h_2c}
} v_{ih_1[i]h_2ck}z_{ih_1[i]h_2ck}} {\sum_{h_2=1}^2 \sum_{c \in S^{\tiny{\mbox{\,AUG}}}_{ih_1[i]h_2}} \sum_{k \in S^{\tiny{\mbox{\,AUG}}}_{ih_1[i]h_2c}}v_{ih_1[i]h_2ck} }.$$
The above derivation provides a rationale for the adjusted estimator which is of the same form as the original Hájek estimator, so that now we can use the same variance calculation as would be used for the original.

For either planned (Admin-1) or unplanned (Admin-2) areas the augmented variance is,
\begin{equation}\label{eq:sasvaraug}
 \widehat{V}(\widehat p_i^{\tiny{\mbox{\,AUG}}}) = 
\frac{1}{(v^{\tiny{\mbox{\,AUG}}}_{ih_1[i]\cdot\cdot\cdot})^2} \sum_{h_2=1}^2 \frac{n^{\tiny{\mbox{\,AUG}}}_{ih_1[i]h_2}}{ n^{\tiny{\mbox{\,AUG}}}_{ih_1[i]h_2}-1}\sum_{c \in S^{\tiny{\mbox{\,AUG}}}_{ih_1[i]h_2}} \left[
v^{\tiny{\mbox{\,AUG}}}_{ih_1[i]h_2c\cdot} ( \widehat p^{\tiny{\mbox{\,AUG}}}_{ih_1[i]h_2c}-\widehat p_i^{\tiny{\,\mbox{AUG}}}) -\frac{1}{ n^{\tiny{\mbox{\,AUG}}}_{ih_1[i]h_2}}
v^{\tiny{\mbox{\,AUG}}}_{ih_1[i]h_2\cdot\cdot}( \widehat p_{ih_1[i]h_2}^{\tiny{\,\mbox{AUG}}}-\widehat p_i^{\tiny{\,\mbox{AUG}}})
\right]^2
\end{equation}

where $ n_{ih_1[i]h_2}^{\tiny{\mbox{\,AUG}}} =  n_{ih_1[i]h_2}+ n^{\tiny{\mbox{\,PH}}}_{ih_1[i]h_2}$, with $ n^{\tiny{\mbox{\,PH}}}_{ih_1[i]h_2}$ as the number of phantom clusters in area $i$, strata $h_1[i],h_2$, and 
$$v^{\tiny{\mbox{\,AUG}}}_{ih_1[i]h_2c\cdot} = \sum_{k \in S^{\tiny{\mbox{\,AUG}}}_{ih_1[i]h_2c}} v_{ih_1[i]h_2ck},\qquad 
v^{\tiny{\mbox{\,AUG}}}_{ih_1[i]h_2\cdot \cdot} = \sum_{c \in S_{ih_1[i]h_2}} v^{\tiny{\mbox{\,AUG}}}_{ih_1[i]h_2c\cdot}
\qquad
v^{\tiny{\mbox{\,AUG}}}_{ih_1[i]\cdot\cdot\cdot} = \sum_{h_2=1}^2
v^{\tiny{\mbox{\,AUG}}}_{ih_1[i]h_2\cdot \cdot}
$$
and
$$ 
\widehat p_{ih_1[i]h_2}^{\tiny{\mbox{\,AUG}}} = \frac{\sum_{c \in S^{\tiny{\mbox{\,AUG}}}_{ih_1[i]h_2}} \sum_{k \in S^{\tiny{\mbox{\,AUG}}}_{ih_1[i]h_2c}} v_{ih_1[i]h_2ck} z_{ih_1[i]h_2ck}}{v^{\tiny{\mbox{\,AUG}}}_{ih_1[i]h_2\cdot\cdot}},\qquad \widehat p_{ih_1[i]h_2c}^{\tiny{\mbox{\,AUG}}} = \frac{\sum_{k \in S^{\tiny{\mbox{\,AUG}}}_{ih_1[i]h_2c}} v_{ih_1[i]h_2ck} z_{ih_1[i]h_2ck}}{v^{\tiny{\mbox{\,AUG}}}_{ih_1[i]h_2c\cdot}}.$$

\subsection{Rules for Variance Fix}\label{sec:rules}


For both planned and unplanned domains, we delineate two cases in which the variance estimator fails and a variance fix procedure is required.

\begin{enumerate}
\item {\it Single  Cluster Strata:} When any stratum defined by $i,h_1[i],h_2$ has only one sampled cluster, i.e.,~$ n_{ih_1[i]h_2} = 1$, the variance formula, (\ref{eq:sasvar}), is undefined. This case rarely occurs with DHS survey data in practice for planned areas since the planned $n_{ih_1[i]h_2}$ is much larger than 1. Recall that in an unplanned domain, we still use $n_{ih_1[i]h_2}$ in (\ref{eq:sasvar}), rather than the number of clusters in the unplanned domain.

Our suggested modification is to add one phantom cluster in each strata within which $n_h = 1$. \\

\item {\it Identical Estimates:} The variance formula also breaks down when $\widehat p_i = \widehat{p}_{ih_1[i]h_2} = \widehat{p}_{ih_1[i]h_2c}$ for all $c$ in the domain (no matter whether it is planned or unplanned). Note here that $\widehat p_i$ is the domain estimate and in the DHS we have $\widehat{p}_{ih_1[i]h_2}$, with $h_2$ corresponding to urban and rural regions of the domain of interest $i$ and  $\widehat{p}_{ih_1[i]h_2c}$ are the cluster estimates. In DHS survey data, this case mainly occurs in unplanned domains.
This scenario occurs if an unplanned domain contains only one sampled cluster, since the domain estimate is then identical to that single cluster’s estimate.
\vspace{.15in}

\end{enumerate}

In this second case, phantom clusters are added only to the strata in which the identical estimates condition holds. For example:
\begin{itemize}
    \item If a problematic Admin-2 area contains sampled clusters only in one of the stratum and satisfies 
    $\widehat{p}_i = \widehat{p}_{ih_1[i]h_2} = \widehat{p}_{ih_1[i]h_2c}$, then one phantom cluster is added to the urban stratum alone.
    \item If both the urban and rural strata satisfy the  identical-estimates condition, then one phantom cluster is added to each stratum.
\end{itemize}
In both cases, each phantom cluster is assigned a prior mean equal to the corresponding national strata specific Hájek estimate (for example, urban or rural) with prior weights equal to the national average survey weights computed across all clusters in the stratum. This prior is in the same spirit as unit information prior \citep{kass1995reference} in which a minimal amount of information is used in the prior, which is therefore weakly data dependent.

In the Zambia case study of unplanned domains (Admin-2 areas), 24 require variance modification:~10 have only one sampled cluster, while the remaining 14  contain multiple clusters but have  identical cluster means, see Table \ref{tab:illegal} for more details. We refer to any failure of the original variance formula as an \textit{illegal} variance case; all other situations are treated as \textit{legal} variance cases.

\section{Simulation Study}\label{sec:simulation}

\subsection{Simulation Design}

In this section we examine the properties of the adjustment we have proposed. 
To mimic our motivating data, we take the geographical areas from Zambia and simulate from a fixed population according to a stratified, two-stage unequal probability cluster design. We slightly simplify the design relative to the DHS, and the notation accordingly, and do not include urban/rural in the stratification so that the strata are the $H=10$ Admin-1 areas (planned domains). 

The targets for inference are the $I=115$ Admin-2 areas, which are unplanned domains. The sampling frame for the DHS was based on the 2010 census, with some updates to accommodate changes in Admin-1 and Admin-2 areas since 2010. 

While \cite{ZambiaDHS:18} provides relevant information at Admin-1 level, key details about the unplanned domains (Admin-2 areas) in the sampling frame, including the population size and the number of clusters, are unavailable. As a result, approximations are required.
For each Admin-2 area, the number of clusters in the frame $C_{ih}$ is approximated based on its share of the total population within its corresponding Admin-1 region, $h=1,\dots,10$, $i=1,\dots,I$. For Admin-1 area $h$, and Admin-2 area $i$, the number of clusters is taken as a population-weighted allocation:
\begin{equation}
    C_{ih} = \text{round} \left( \frac{N_{ih}}{N_{+h}} \times C_{+h} \right)
\end{equation}
where:
\begin{itemize}
    \item $N_{ih}$ is the population of the $i$-th Admin-2 area in Admin-1 area $h$, based on WorldPop \citep{tatem2017worldpop},
    \item $N_{+h}$ is the total population of Admin-1 area $h$,
    \item $C_{+h}$ is the total number of clusters in Admin-1 area $h$, based on the DHS report \citep[Table A.2]{ZambiaDHS:18}. There are $C_{++}=25,631$ clusters in the sampling frame.   
\end{itemize}

For cluster $c$, we generate a population size $N_{ihc}$, $c=1,\dots,C_{ih}$
by proportionally allocating 
the total Admin-2 population $N_{ih}$ according to normalized random fractions 
drawn from an Exponential(1) 
distribution, while enforcing a minimum cluster size of 
30 individuals. The binary outcome for individual $k$ in cluster $c$ is drawn from a Bernoulli distribution with probability $p_{ck}$ with:
$$
    \text{logit}(p_{ck}) = \text{logit}(m_0) + \alpha_{i[c]} + e_c + e_{ck}
$$
where:
\begin{itemize}
    \item $m_0$ is the baseline prevalence rate at the national level,
    \item $\alpha_{i[c]} \sim \mbox{N}(0, 0.5^2)$ is an area-level random effect (at the Admin-2 level), where $i[c]$ represents the Admin-2 area $i$ within which the $c$-th cluster is contained, 
    \item $e_c \sim \mbox{N}(0,  0.2^2)$ is a cluster-level random effect,
    \item $e_{ck} \sim \mbox{N}(0,  0.05^2)$ is an individual-level random effect.
\end{itemize}
We take wasting prevalence (with DHS code  {\tt CN\_NUTS\_C\_WH2}) in Zambia as the indicator from which simulation settings are constructed. Specifically, the prevalence of wasting in under 5 children (standardized weight for height, WHZ $< -2$) has a national  prevalence of 0.042 \citep[p.~xxiii]{ZambiaDHS:18}.

Within strata (Admin-1 area) $h$, we perform probability proportional to size (PPS) sampling of clusters. We base the selection probability for cluster $c$ as proportional to its population size (in the Zambia DHS, the size variable was taken as the number of households, but we simplify), and the number of clusters selected follows the DHS design for that stratum, using PPS,
\begin{equation}
    \pi_{ihc}^{(1)} = n_h  \times \frac{
    N_{ihc}}
    {N_{+h}}  
\end{equation}

where $n_h$ is the intended sample size for strata $h$, taken from the DHS report \citep[Table A.3]{ZambiaDHS:18}.
At stage 2, for each Admin-2 area we
sample 30 individuals per cluster, as in  the DHS design. 
\begin{equation}
    \pi_{ihc}^{(2)} = \frac{30}{N_{ihc}}.
\end{equation}
The design weight for cluster $c$ in stratum $h$ is
$w_{hic} = 1/(\pi_{ihc}^{(1)} \times \pi_{ihc}^{(2)})$.

Three variance-handling strategies were compared to evaluate the estimation methods in each simulation:
\begin{enumerate}
    \item {\it All-Unfixed:} 
Use the variance estimate in \eqref{eq:sasvar} for all domains.
\item {\it All-Fixed:} 
Use variance adjustment formula \eqref{eq:sasvaraug} for all domains, 
regardless of whether the estimated variance is legal or not.
\item {\it Fixed (Illegal) and Unfixed (Legal):} 
Use variance estimate \eqref{eq:sasvaraug} only for domains which have illegal variances, with \eqref{eq:sasvar} used for legal variances. 
\end{enumerate}

\subsection{Simulation Metrics}

We evaluate several key metrics to assess model performance. 
All metrics are computed for each unplanned domain (i.e.,~Admin-2 area) and for each estimation method. Recall $i =1,\dots,I$ indexes Admin-2 areas and $m=1,\dots,M$ methods (with $M=3$). 
We simulate $S = 1000$ datasets, but because some Admin-2 areas may not contain clusters in every simulation (they are unplanned domains), we let $S_{i}$  denote the number of simulated datasets (out of $S$) for which domain 
$i$ contains at least one sampled cluster, so that $S_i \leq S$.

The true population prevalence in unplanned domain $i$ is $p_{i}$ and for simulation $s =1,\dots,S$, in $i$, $i=1,\dots,I$, define the following quantities:
\begin{itemize}
  \item $\widehat{p}^{\;\tiny{\text{w}(s)} }_{im}$: method $m$ prevalence estimate,
  \item $\widehat \theta_{im}^{\tiny{\;\text{w}(s)} }= \mbox{logit}(\widehat{p}^{\tiny{\;\text{w}(s)} }_{im})$: method $m$ logit prevalence estimate,
  \item $\widehat V^{(s)}_{im}$: method $m$ estimated variance of logit-prevalence estimate.
\end{itemize}

The asymptotic 100 $\times (1-\alpha)\%$ confidence interval for simulation $s$ is:
$$
\begin{aligned}
L^{(s)}_{im} &= \mathrm{expit}\left( ~\widehat \theta_{im}^{\tiny{\;\text{w}(s)} } - z_{1-\alpha/2} \times \sqrt{\widehat V^{(s)}_{im}} ~\right), &
U^{(s)}_{im} &= \mathrm{expit}\left( ~\widehat \theta_{im}^{\tiny{\;\text{w}(s)} }+ z_{1-\alpha/2} \times \sqrt{\widehat V^{(s)}_{im}} ~\right), \\
\end{aligned}
$$
where  $z_{1-\alpha/2}$ is the quantile of the standard normal distribution.

Note that if the variance is illegal and no fix is applied, we take the upper and lower bounds to equal the point estimate $\widehat{p}^{\tiny{\;\text{w}(s)}}_{im}$. The variance for the unplanned domain is defined as illegal
following the rules defined in Section \ref{sec:rules}.

For each metric, we compute its value for every area and simulation and average over all valid simulations $s =1,\dots,S_{i}$. 
\vspace{.2in}

\noindent
{\bf Coverage:} Coverage measures the proportion of simulations in which the true value for domain $i$, $p_{i}$, lies within the constructed interval:  
$$
\mathrm{Coverage}_{im}
= \frac{1}{S_{i}} \sum_{s = 1}^{S_{i}} 
I\left(L^{(s)}_{im} \le p_{i} \le U^{(s)}_{im}\right).
$$

\vspace{.2in}

\noindent
{\bf Confidence Interval Width:} The width of the confidence interval quantifies the average uncertainty range across simulations:
\[
\mathrm{Width}_{im}
= \frac{1}{S_{i}} \sum_{s = 1}^{S_{i}}  \left (U^{(s)}_{im} - L^{(s)}_{im}\right).
\]
When the variance is illegal, the width is 0.
\vspace{.2in}

\noindent
{\bf Interval Score:} The interval score balances interval width with coverage.  
The mean interval score is,
$$
\mathrm{IS}_{im} 
= \frac{1}{S_{i}} \sum_{s=1}^{S_{i}}   \left\{ \left(U^{(s)}_{im} - L^{(s)}_{im} \right)
+ \frac{2}{\alpha}
\left[
  \left(p_i - U^{(s)}_{im}\right)_+
  +
  \left(L^{(s)}_{im} - p_i\right)_+
\right] \right\},
$$
where $(x)_+ = \max(x, 0)$ denotes the positive part. 
This score is intuitive since it rewards narrow intervals that have good coverage
\citep{gneiting2014probabilistic} so 
lower interval scores indicate better performance.  We take $\alpha=0.2$ for all metrics.

\subsection{Simulation Results}

\begin{figure}[H]
    \centering
    \includegraphics[width=\textwidth]{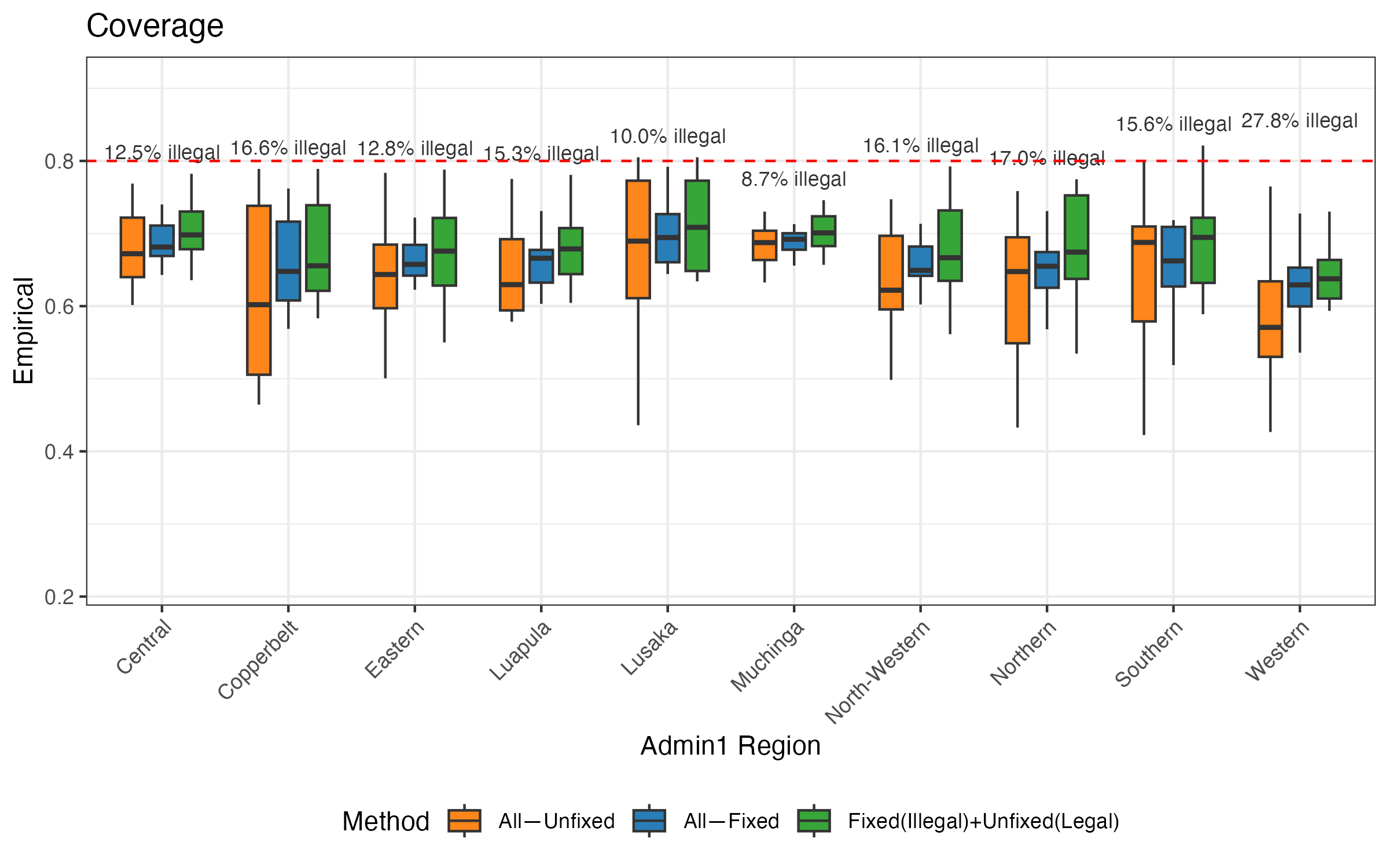}
    \caption{Empirical coverage across Admin-2 (unplanned) areas, under the nominal 80\% level (red dashed line), using asymptotic   
        normal sampling distribution of the estimator.  
        Annotations show, within each planned domain, the percentage of unplanned domains with illegal direct variance estimates.
    }
    \label{fig:coverage_0.8}
\end{figure}

In Figure \ref{fig:coverage_0.8}, we display the coverages (with 80\% nominal) from the simulation. All three of the methods display undercoverage, but the best performing method is that which modifies only those areas that require fixing. The poorest method is that which does not fix any of the variances; this method also produces the largest variation across areas. The narrowest variation is with the all fixed method.  The proportion of Admin-2 areas with illegal variances is added to the figure and across Admin-1 areas the ranges is 8.7\%--27.8\%.  The CI width plot is included in Appendix C.2 of the Supplementary Materials.
In Appendix C.1 of the supplementary Materials we present additional simulations in which we increase the sample sizes and see that the nominal coverage is recovered.

\begin{figure}[H]
    \centering
    \includegraphics[width=0.8\textwidth]{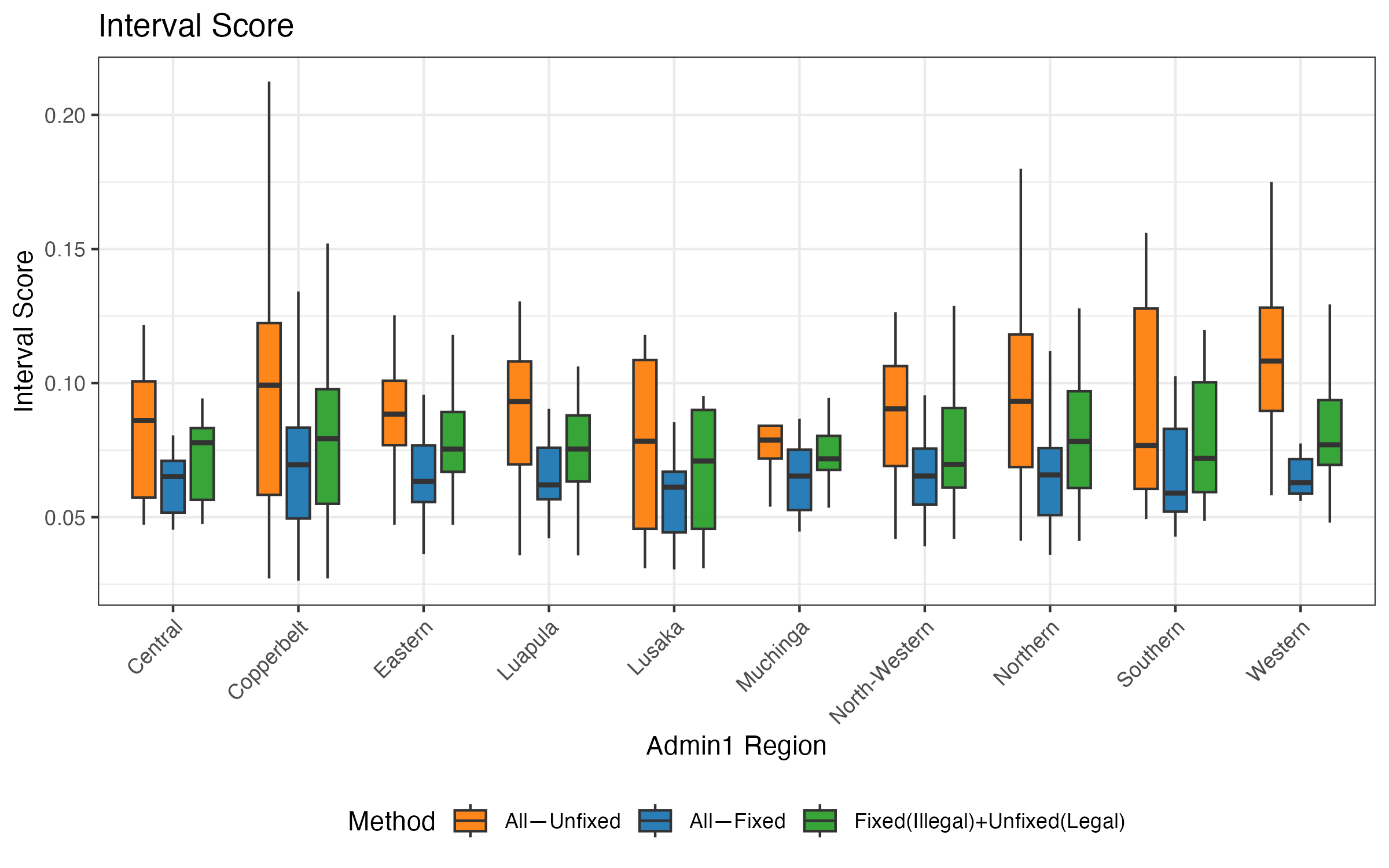}
    \caption{Interval scores (with 80\% coverage) for Admin-2 domains for the three 
        variance-handling strategies, by Admin-1 domains.
    }
    \label{fig:ISs}
\end{figure}

In Figure \ref{fig:ISs} we display the interval scores for Admin-2 areas and see that the lowest (preferable) scores are always from the method in which all areas are fixed, followed by the method that modifies only those areas that need fixing, with the all unfixed method giving the largest scores.

\section{Childhood Wasting in Zambia}\label{sec:zambia}

Based on direct estimates, 
(see Figure~\ref{fig:Zambia_map} for province names), Muchinga province
has the highest childhood wasting among Admin-1
areas, with a prevalence of 0.082 (0.058--0.115), which is more than 3.7
times greater than Eastern province, with prevalence 0.022 (0.014--0.035), which has lowest point estimate. 
 At Admin-2 level, 3 regions have no data and 24 do not produce a variance for the direct estimates, and so require modification. Of these, 10 have only one sampled cluster, while the remaining 14 contain multiple clusters but have identical cluster means, see Table \ref{tab:illegal} for details.
 
To estimate wasting prevalence across districts (Admin-2 areas) of Zambia we examine four methods:
\begin{enumerate}
    \item Unadjusted direct estimates. Only 88  from 115 areas produce estimates and standard errors under this approach.
    \item Adjusted direct estimates use phantom clusters to modify the prevalences and variances.
    \item Unadjusted Fay-Herriot, which treat as missing data and predict the prevalence for those 27 areas with no data or problematic variances. The spatial model aids greatly in this endeavor.
    \item Adjusted Fay-Herriot use phantom clusters in 24 areas, to modify the prevalences and variances. The 3 areas with no data are treated as missing values and are predicted from the model. 
\end{enumerate} 

When Admin-2 estimates are sought, and random effects are introduced at Admin-2 , it is possible for overshrinkage to occur, particularly for rare outcomes.
To provide some protection against this, for both of Fay-Herriot models, we use {\it nested spatial models}. Specifically, we replace (\ref{eq:fayherriot2}) with linking model,
$$\theta_i = \alpha_{h_1[i]} + u_i,$$
where $h_1[i]$ is the index of the Admin-1 area within which Admin-2 area $i$ is nested. The parameters $\alpha_{h_1}$, $h_1=1,\dots,10$, are treated as fixed effects, and the $u_i$ are taken as BYM2 random effects, via equation (\ref{eq:fayherriotBYM2}), to model {\it within Admin-1 variation}. We follow a Bayesian approach with PC priors, and use the 
 integrated nested Laplace approximation (INLA) for computation \citep{rue:etal:17}. INLA is ideal for computation for hierarchical, and more specifically, spatial models, and is extremely fast and accurate for such models. For the models we fit to Zambia, the prior adjustment and fitting of the nested Fay-Herriot model takes $<1$ minute. We measured wall-clock fit times on a desktop with an Intel Core i7-8700K (6 physical cores/12 threads, base 3.70 GHz), using 1 thread.

\begin{figure}[]
    \centering
    \includegraphics[height=22cm]{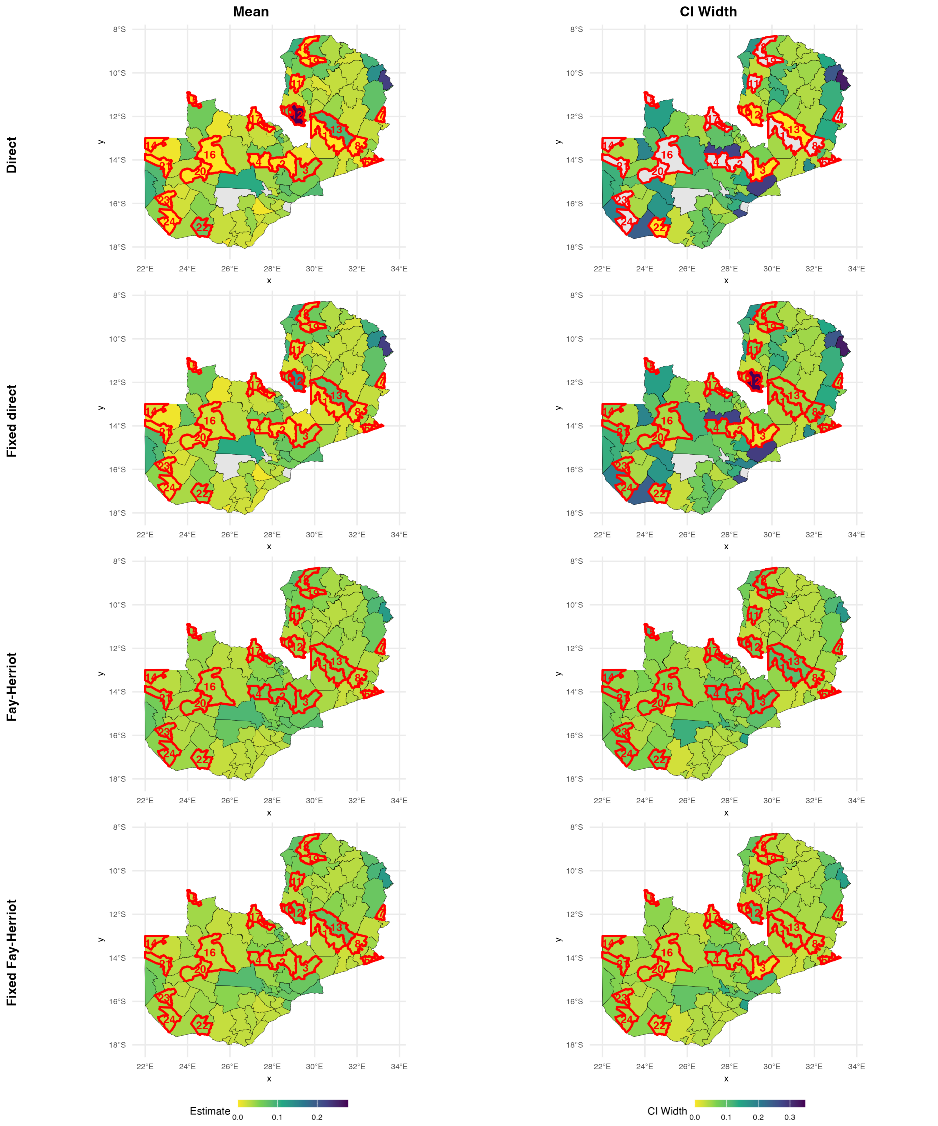}
    \caption{Prevalence estimates and 95\% uncertainty interval width for four approaches.
        The red borders and numbers indicate areas with illegal variance estimates. Information on these areas is presented in Table \ref{tab:illegal}.  In the left column, and in the second row of the right column, areas in gray have no clusters. In the top right map, the gray areas have no clusters or undefined/zero variance estimates.
    }
    \label{fig:combined_map}
\end{figure}

Figure \ref{fig:combined_map} presents maps of point estimates (left) and confidence/credible interval width (right). The greater variation in the Admin-2 direct estimates is clear, as is the reduction of CI widths under the Fay-Herriot models. Appendix D gives further results that compare non-nested and nested Fay-Herriot models, and  point estimates and uncertainty measures under different approaches. We also summarize the hyperparameters for nested and non-nested with modified and non-modified variances. As expected, the spread of the BYM2 random effects is narrower under the nested models, since some of the spatial variation is incorporated into the fixed effects. In the modified variance models the proportion of variation that is spatial is smaller than the non-modified version that treats areas with problematic variances as missing. One interpretation is that the variance modification is acting as a form of pre-smoothing, but this needs further investigation.

\begin{table}[htbp]
\centering
\caption{Information on the Admin-2 areas that need modification. The Index numbers correspond to the labels in Figure \ref{fig:combined_map}.
}\label{tab:illegal}
\fontsize{8}{10}\selectfont
\scalebox{0.85}{
\begin{tabular}{rlllrrrrr}
\toprule
Index & Admin-2 & Admin-1 & Urban/Rural & ClusterID & $n_{\text{trials}}$ & outcome & phantom mean & phantom weights\\
\midrule
1 & Chitambo      & Central     & rural & 5   & 32 & 0 & 0.038 & 18063987\\
1 & Chitambo      & Central     & rural & 60  & 22 & 0 & 0.038 & 18063987\\
\addlinespace
2 & Kapiri Mposhi & Central     & rural & 24  & 23 & 0 & 0.038 & 18063987\\
2 & Kapiri Mposhi & Central     & rural & 136 & 22 & 0 & 0.038 & 18063987\\
2 & Kapiri Mposhi & Central     & rural & 167 & 26 & 0 & 0.038 & 18063987\\
2 & Kapiri Mposhi & Central     & rural & 255 & 24 & 0 & 0.038 & 18063987\\
2 & Kapiri Mposhi & Central     & rural & 308 & 14 & 0 & 0.038 & 18063987\\
2 & Kapiri Mposhi & Central     & rural & 373 & 12 & 0 & 0.038 & 18063987\\
2 & Kapiri Mposhi & Central     & rural & 476 & 17 & 0 & 0.038 & 18063987\\
2 & Kapiri Mposhi & Central     & urban & 302 & 16 & 0 & 0.049 & 16662007\\
2 & Kapiri Mposhi & Central     & urban & 508 & 6  & 0 & 0.049 & 16662007\\
\addlinespace
3 & Luano         & Central     & rural & 166 & 22 & 1 & 0.038 & 18063987\\
\addlinespace
4 & Ngabwe        & Central     & rural & 247 & 15 & 0 & 0.038 & 18063987\\
\addlinespace
5 & Chingola      & Copperbelt  & rural & 297 & 22 & 0 & 0.038 & 18063987\\
5 & Chingola      & Copperbelt  & rural & 379 & 15 & 0 & 0.038 & 18063987\\
5 & Chingola      & Copperbelt  & urban & 116 & 23 & 0 & 0.049 & 16662007\\
5 & Chingola      & Copperbelt  & urban & 214 & 11 & 0 & 0.049 & 16662007\\
5 & Chingola      & Copperbelt  & urban & 368 & 8  & 0 & 0.049 & 16662007\\
5 & Chingola      & Copperbelt  & urban & 392 & 8  & 0 & 0.049 & 16662007\\
5 & Chingola      & Copperbelt  & urban & 427 & 15 & 0 & 0.049 & 16662007\\
\addlinespace
6 & Chadiza       & Eastern     & rural & 333 & 7  & 0 & 0.038 & 18063987\\
\addlinespace
7 & Chasefu       & Eastern     & rural & 221 & 21 & 0 & 0.038 & 18063987\\
7 & Chasefu       & Eastern     & rural & 310 & 17 & 0 & 0.038 & 18063987\\
7 & Chasefu       & Eastern     & rural & 358 & 14 & 0 & 0.038 & 18063987\\
7 & Chasefu       & Eastern     & rural & 422 & 19 & 0 & 0.038 & 18063987\\
\addlinespace
8 & Mambwe        & Eastern     & rural & 82  & 9  & 0 & 0.038 & 18063987\\
8 & Mambwe        & Eastern     & rural & 280 & 19 & 0 & 0.038 & 18063987\\
\addlinespace
9 & Vubwi         & Eastern     & rural & 25  & 27 & 0 & 0.038 & 18063987\\
9 & Vubwi         & Eastern     & rural & 527 & 16 & 0 & 0.038 & 18063987\\
\addlinespace
10 & Chembe       & Luapula     & rural & 451 & 28 & 2 & 0.038 & 18063987\\
\addlinespace
11 & Chipili      & Luapula     & rural & 160 & 25 & 0 & 0.038 & 18063987\\
\addlinespace
12 & Milengi      & Luapula     & rural & 36  & 18 & 5 & 0.038 & 18063987\\
\addlinespace
13 & Lavushimanda & Muchinga    & rural & 11  & 22 & 2 & 0.038 & 18063987\\
13 & Lavushimanda & Muchinga    & rural & 482 & 33 & 3 & 0.038 & 18063987\\
\addlinespace
14 & Chavuma      & North-Western & rural & 30  & 17 & 0 & 0.038 & 18063987\\
\addlinespace
15 & Ikelenge     & North-Western & rural & 73  & 16 & 0 & 0.038 & 18063987\\
15 & Ikelenge     & North-Western & rural & 376 & 20 & 0 & 0.038 & 18063987\\
\addlinespace
16 & Mufumbwe     & North-Western & rural & 27  & 22 & 0 & 0.038 & 18063987\\
16 & Mufumbwe     & North-Western & rural & 120 & 16 & 0 & 0.038 & 18063987\\
16 & Mufumbwe     & North-Western & urban & 123 & 22 & 0 & 0.049 & 16662007\\
\addlinespace
17 & Mushindano   & North-Western & rural & 481 & 21 & 0 & 0.038 & 18063987\\
\addlinespace
18 & Kaputa       & Northern    & rural & 39  & 20 & 0 & 0.038 & 18063987\\
18 & Kaputa       & Northern    & rural & 459 & 19 & 0 & 0.038 & 18063987\\
18 & Kaputa       & Northern    & urban & 505 & 16 & 0 & 0.049 & 16662007\\
\addlinespace
19 & Mporokoso    & Northern    & rural & 259 & 17 & 0 & 0.038 & 18063987\\
19 & Mporokoso    & Northern    & urban & 216 & 10 & 0 & 0.049 & 16662007\\
19 & Mporokoso    & Northern    & urban & 313 & 18 & 0 & 0.049 & 16662007\\
\addlinespace
20 & Kaoma        & Western     & rural & 103 & 20 & 0 & 0.038 & 18063987\\
20 & Kaoma        & Western     & rural & 367 & 22 & 0 & 0.038 & 18063987\\
20 & Kaoma        & Western     & rural & 473 & 15 & 0 & 0.038 & 18063987\\
20 & Kaoma        & Western     & urban & 415 & 14 & 0 & 0.049 & 16662007\\
20 & Kaoma        & Western     & urban & 421 & 16 & 0 & 0.049 & 16662007\\
\addlinespace
21 & Mitete       & Western     & rural & 499 & 30 & 0 & 0.038 & 18063987\\
\addlinespace
22 & Mwandi       & Western     & rural & 463 & 14 & 1 & 0.038 & 18063987\\
\addlinespace
23 & Nalolo       & Western     & rural & 218 & 10 & 0 & 0.038 & 18063987\\
23 & Nalolo       & Western     & rural & 299 & 12 & 0 & 0.038 & 18063987\\
23 & Nalolo       & Western     & rural & 362 & 13 & 0 & 0.038 & 18063987\\
\addlinespace
24 & Sioma        & Western     & rural & 140 & 19 & 0 & 0.038 & 18063987\\
24 & Sioma        & Western     & rural & 187 & 11 & 0 & 0.038 & 18063987\\
\bottomrule
\end{tabular}
}
\end{table}

\begin{figure}[]
    \centering
    \includegraphics[height=0.93\textheight]{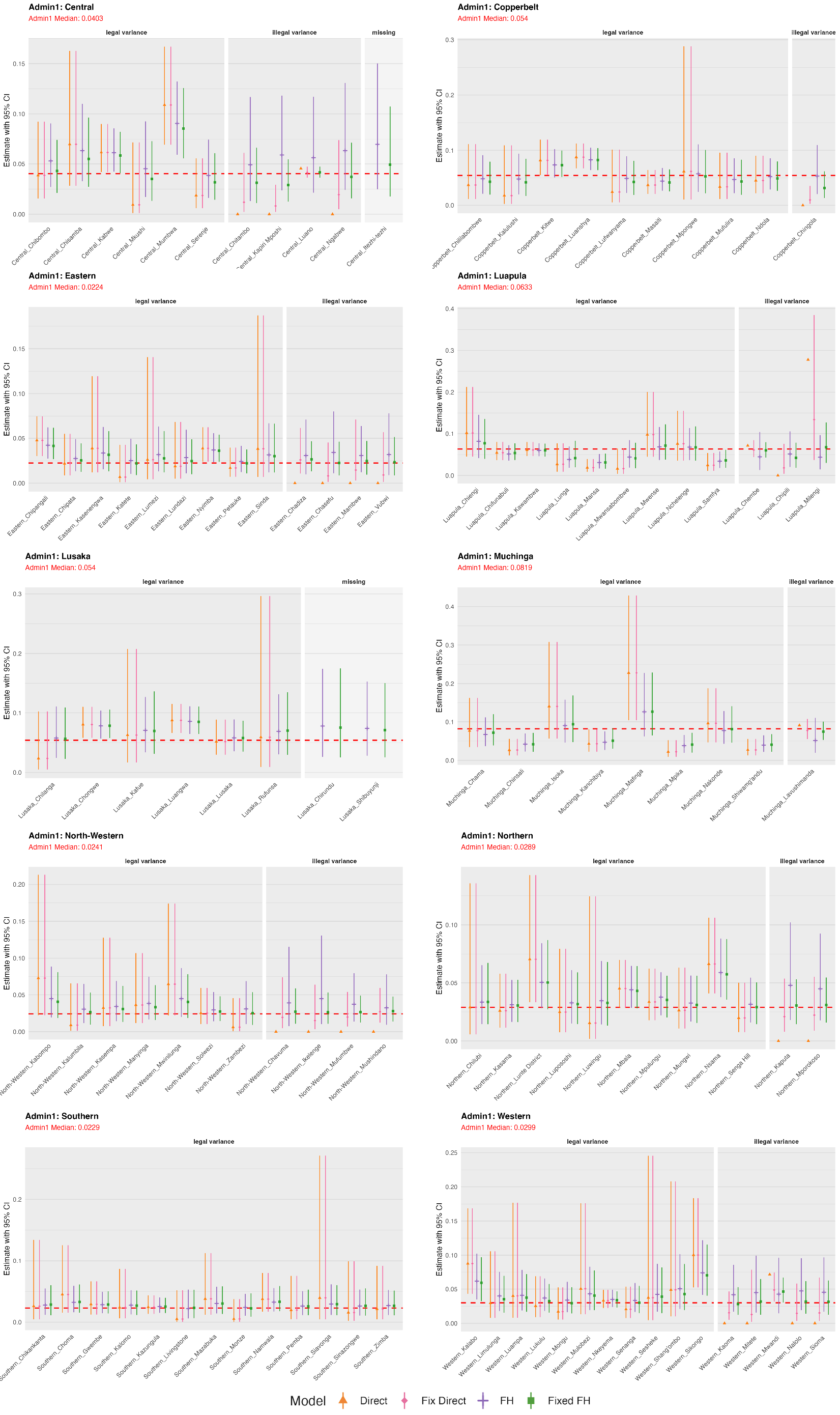}
\caption{Admin-2 estimates by Admin-1 strata, under different models. Vertical lines are 95\% intervals with posterior medians indicated as short horizontal bars. The red dashed line corresponds to the posterior median of the Admin-1 estimate.}
    \label{fig:combined_interval}
\end{figure}

In Figure \ref{fig:combined_interval} we summarize results across Admin-2 areas.
On the right hand side of each panel we plot the summaries for areas with no data, or areas with illegal variances. In general, the variances of the latter are narrower than when treated as missing (as is done in the unmodified Fay-Herriot model).

\begin{figure}[]
    \centering
    \includegraphics[width=1\textwidth]{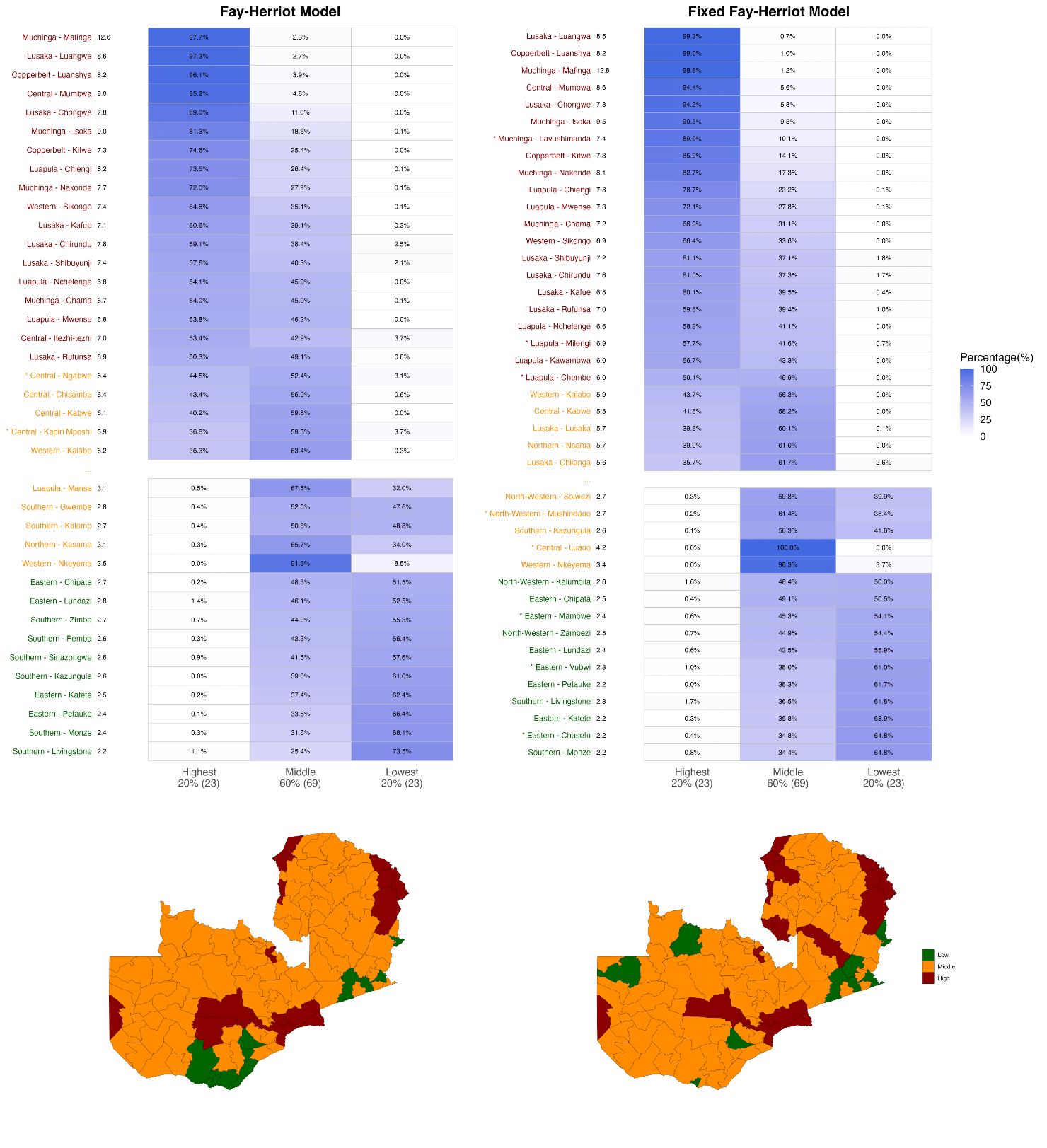}
    \caption{Rankings under original (left) and modified Fay-Herriot models. We tabulate posterior probabilities of Admin-2 districts lying in the highest 20\%, middle 60\%, lowest 20\% (which corresponds to 23, 69, 23 out of 115 areas). The areas with posterior probabilities greater than 50\% of lying in the high prevalence group are highlighted in red, those posterior probabilities greater than 50\% of lying in the low prevalence group are in green, and the remainder are in orange (to save space we do not tabulate all orange areas, which are of less interest). Numbers in the second column of the two plots are the prevalence point estimates.
        Area with illegal variance have a $^\star$ in front of their names.
    }
    \label{fig:ranking}
\end{figure}

In our experience, public health officials are often primarily interested in ranking areas, particularly Admin-2 areas by which health care resources are often administered. Rankings via the prevalence point estimates alone is problematic as it does not account for uncertainty, which is represented by the posterior distribution in our Bayesian Fay-Herriot models. We have posterior samples and for each of these a rank may be applied, e.g.,~1 for the area with the highest prevalence, 2 for that with the second highest, etc. Over all samples and for each rank we have a probability distribution over the areas.  We use these ranking distributions to calculate the posterior probability that each area falls within (say) the highest 20\% of areas, the middle 60\% of areas and the lowest 20\% of areas. In  Figure \ref{fig:ranking}, we give a table of the 20/60/20 breakdown of Admin-2 areas. Areas whose names are in red, have at least a 50\% posterior probability of being in the 20\% of areas with highest prevalence. Areas whose names are in green, have at least a 50\% posterior probability of being in the 20\% of areas with lowest prevalence. Areas in orange are in the middle and we present only a truncated list, since these areas are the least interesting in terms of being high/low. We present the rankings for the unmodified and modified Fay-Herriot models. For the former, the highest prevalence point estimate is for Mafinga district in the province of Muchinga, but it does not have the hightest posterior probability of being in the highest 20\% of districts.

The Lavushimanda District of Muchinga Province (area 13 in Figure \ref{fig:combined_map}) provides an interesting case study. In this area, 2 rural clusters and 0 urban clusters were sampled. In the 2 clusters, 2 from 22 children and 3 children from 33 were wasted. Hence, the prevalence estimate in each cluster is 0.0909 (since all weights in a clusters are equal, the prevalence estimate is the simple average), which is also the rural prevalence estimate, and the province estimate. Since all estimates are equal, the variance estimate is zero also. In Figure \ref{fig:combined_interval} we see that the unmodified Fay-Herriot prevalence is low, because this estimate is based on the neighbors, which all have lower point estimates than Lavushimanda (Figure \ref{fig:combined_map}). The modified Fay-Herriot is relatively large and this leads to it having a prevalence of 0.074 (0.054--	
0.100), compared with 0.051 (0.020--0.110) under the original Fay-Herriot model. In Figure \ref{fig:ranking} we see that Lavushimanda is ranked at 7 with a 89.9\% chance of falling in the 20\% of Admin-2 areas with the highest prevalence, whereas under the original model it appears at rank 35 with a prevalence 0.051 (0.020--0.110) and only a 22\% chance of falling in the highest 20\%. Given the raw prevalence estimate in the 2 clusters, a high ranking is eminently plausible.

\section{Discussion}\label{sec:discussion}

In this paper we have described a variance modification that is principled, retains the design consistency of Fay-Herriot models, and may be automatically applied when one wishes to fit Fay-Herriot models but there are issues with variance calculation in some of the areas. Modeling the variance, via  generalized variance functions \citep[Chapter 7]{wolter:07} or by jointly modeling the mean and variance \citep{gao2023spatial}, provides an alternative approach. However, this endeavor is more sophisticated and consequently difficult to automate. We work with National Statistics Offices in LMICs and typically the need is for multiple indicators to be examined, and it would be very time-consuming and infeasible to use complex variance models. In the LMICs context, reliable covariates to use in prevalence and variance models are hard to come back, in part because censuses are rarely every 10 years and often inaccurate. 
The modification is available in the {\tt surveyPrev} package in {\tt R}\footnote{\url{https://cran.r-project.org/web/packages/surveyPrev/index.html}}.
An important, yet open, question is to give guidelines on what proportion of areas it is reasonable to modify, and this is something we plan to investigate.

Unit-level models provide an alternative approach. For example, zero counts present no problems for binomial models, though overshrinkage with sparse data is always a concern. Further, for prevalence estimation these models are nonlinear, and require a complex aggregation step, and do not give design consistent inference. For examples of unit-level overdispersed binomial models, see \cite{wakefield25two}.

We have presented simulations and a real application that used stratified two-stage cluster sampling, but the method can be applied to general designs. If a variance formula is available, the data may be supplemented with the phantom clusters and the same formula can be used. An alternative approach is to use resampling techniques, such as the jackknife \citep{pedersen:liu:12}.

\section*{Acknowledgments}

This manuscript has been authored by UT-Battelle, LLC, under contract DE-AC05-00OR22725 with the US Department of Energy (DOE). The US government retains and the publisher, by accepting the article for publication, acknowledges that the US government retains a nonexclusive, paid-up, irrevocable, worldwide license to publish or reproduce the published form of this manuscript, or allow others to do so, for US government purposes. DOE will provide public access to these results of federally sponsored research in accordance with the DOE Public Access Plan (https://www.energy.gov/doe-public-access-plan).

\bibliographystyle{natbib} 
\bibliography{spatepi2}

\newpage 
\setcounter{section}{0}
\setcounter{figure}{0}
\setcounter{table}{0}
\setcounter{equation}{0}
\setcounter{page}{1}

\makeatletter
\renewcommand \thesection{S\@arabic\c@section}
\renewcommand \thefigure{S\@arabic\c@figure}
\renewcommand \thetable{S\@arabic\c@table}
\renewcommand \theequation{S\@arabic\c@equation}
\renewcommand \thepage{S\@arabic\c@page}
\makeatother

\centerline{{\huge Supplemental Materials for ``Automatic Variance }}
\vspace{3pt}
\centerline{{\huge Adjustment for Small Area Estimation"}}
\vspace{10pt}

\section{Variance Formulas}

We focus on stratified two-stage cluster sampling, as this is a common design, and provides the main application within the paper.
We need to consider the urban/rural stratification of DHS, since the target areas generally contain both urban and rural segments. 

Before giving the form of the prevalence estimator and its variance we define some notation that will allow us to give expressions for both planned and unplanned domains, indexed by $i$ with $U_i$ representing the population units in area $i$.
We will let $h_1=1,\dots,H_1$ represent the Admin-1 areas that form one element of the stratification and let $h_2=1,2$ represent the second urban/rural element. 
Let $y_{h_1h_2ck}=0/1$ be the binary response, and $w_{h_1h_2ck}$ be the associated design weight, for sampled unit $k$ in cluster $c$ of the stratum determined by $h_1$ crossed with $h_2$,  $c \in S_{h_1h_2}$,  $k \in S_{h_1h_2c}$, for $h_1=1,\dots,H_1$, $h_2=1,2$. 
Hence, $S_{h_1h_2}$ are the set of sampled clusters in the stratum defined by $h_1$ and $h_2$ and $S_{h_1h_2c}$ are the set of individuals who are subsequently selected in cluster $c$. 
 
Following standard practice \citep{sas2016sas}, we define {\it extended domain variables},
\begin{eqnarray*}
    z_{ih_1h_2ck} &=& \left\{ \begin{array}{ll}
      I( k \in U_i ) y_{h_1[i]h_2ck}  &  \mbox{ i.e.,~if unit $k$ belongs to target domain $i$ 
      } 
      \\
      0   & \mbox{ otherwise. }
    \end{array}
    \right.\\
       v_{ih_1h_2ck} &=& \left\{ \begin{array}{ll}
      I( k \in U_i ) w_{h_1[i]h_2ck}  &  \mbox{ i.e.,~if unit $k$ belongs to target domain $i$ 
      } 
      \\
      0   & \mbox{ otherwise. }
    \end{array}
    \right. ,
\end{eqnarray*}
where $I(A)$ is 1 if the event $A$ occurs, and is 0 otherwise.  For both planned and unplanned domains, area $i$ corresponds to exactly one Admin-1 area, $h_1[i]$, with $i=h_1[i]$ for planned domains.

We repeat the form that is given in the paper. The variance of the estimator given in equation (7) of the main paper is \citep[p.~9282--9283]{sas2016sas},
$$ \widehat{V}(\widehat p_i^{\tiny{\,\mbox{HJ}}}) =
\sum_{h_2=1}^2 \widehat{V}(\widehat p_{ih_1[i]h_2}^{\tiny{\,\mbox{HJ}}}),$$
where,
\begin{equation}\label{eq:sasvar2}
 \widehat{V}(\widehat p_{ih_1[i]h_2}^{\tiny{\,\mbox{HJ}}}) = 
\frac{1}{v_{ih_1[i]\cdot\cdot\cdot}^2} \frac{n_{ih_1[i]h_2}}{ n_{ih_1[i]h_2}-1}\sum_{c \in S_{ih_1[i]h_2}} \left[
v_{ih_1[i]h_2c\cdot} ( \widehat p_{ih_1[i]h_2c}^{\tiny{\,\mbox{HJ}}}-\widehat p_i^{\tiny{\,\mbox{HJ}}}) -\frac{1}{n_{ih_1[i]h_2}}
v_{ih_1[i]h_2\cdot\cdot}( \widehat p_{ih_1[i]h_2}^{\tiny{\,\mbox{HJ}}}-\widehat p_i^{\tiny{\,\mbox{HJ}}})
\right]^2
\end{equation}
where $n_{h_1[i]h_2}$ is the number of clusters in strata $h_1[i],h_2$ and
$$
v_{ih_1[i]h_2c\cdot} = \sum_{k \in S_{ih_1[i]h_2c}} v_{ih_1[i]h_2ck},
\qquad 
v_{ih_1[i]h_2\cdot \cdot} = \sum_{c \in S_{ih_1h_2}}v_{ih_1[i]h_2c\cdot},\qquad
v_{ih_1[i]\cdot\cdot\cdot}= \sum_{h_2=1}^2 v_{ih_1[i]h_2\cdot \cdot}
$$
and
$$ 
\widehat p_{ih_1[i]h_2}^{\tiny{\mbox{\,HJ}}} = \frac{\sum_{c \in S_{ih_1[i]h_2}} \sum_{k \in S_{ih_1[i]h_2c}} v_{ih_1[i]h_2ck} z_{ih_1[i]h_2ck}}{v_{ih_1[i]h_2\cdot\cdot}},\qquad \widehat p_{ih_1[i]h_2c}^{\tiny{\mbox{\,HJ}}} = \frac{\sum_{k \in S_{ih_1[i]h_2c}} v_{ih_1[i]h_2ck} z_{ih_1[i]h_2ck}}{v_{ih_1[i]h_2c\cdot}}.$$

For unplanned domains, and in strata $h_1[i],h_2$, let $S_{ih_1[i]h_2}^{\tiny{\mbox{IN}}}$ be the set of clusters that are contained within the target area and $S_{ih_1[i]h_2}^{\tiny{\mbox{OUT}}}$ be the set of clusters that are not contained within the target area but are within the larger planned domain $h_1[i]$. Note that $S_{ih_1[i]h_2} = S_{ih_1[i]h_2}^{\tiny{\mbox{IN}}} \cup S_{ih_1[i]h_2}^{\tiny{\mbox{OUT}}}$
Then, for unplanned domains the variance form \eqref{eq:sasvar2} can be written as,
\begin{eqnarray}
 \widehat{V}(\widehat p_i^{\tiny{\,\mbox{HJ}}}) &=& 
\frac{1}{v_{ih_1[i]\cdot\cdot\cdot}^2}
\sum_{h_2=1}^2 
\frac{n_{ih_1[i]h_2}}{n_{ih_1[i]h_2}-1}
\left(
A_{ih_1[i]h_2} + B_{ih_1[i]h_2}
\right)
\label{eq:sasvar3}
\end{eqnarray}
where
\begin{eqnarray*}
    A_{ih_1[i]h_2}&=& \sum_{c \in S_{ih_1[i]h_2}^{\tiny{\mbox{IN}}}}
\left[
v_{ih_1[i]h_2c\cdot}
( \widehat p_{ih_1[i]h_2c}^{\tiny{\,\mbox{HJ}}}
-\widehat p_i^{\tiny{\,\mbox{HJ}}})
-\frac{1}{n_{ih_1[i]h_2}}
v_{ih_1[i]h_2\cdot\cdot}
( \widehat p_{ih_1[i]h_2}^{\tiny{\,\mbox{HJ}}}
-\widehat p_i^{\tiny{\,\mbox{HJ}}})
\right]^2\\
B_{ih_1[i]h_2} &=&\sum_{c \in S_{ih_1[i]h_2}^{\tiny{\mbox{OUT}}}}
\frac{1}{n_{ih_1[i]h_2}^2}
v_{ih_1[i]h_2\cdot\cdot}^2
( \widehat p_{ih_1[i]h_2}^{\tiny{\,\mbox{HJ}}}
-\widehat p_i^{\tiny{\,\mbox{HJ}}})^2
\end{eqnarray*}
with the $B_{ih_1[i]h_2} $ term inflating the variance to account for the randomness of the number of observations that fall in the domain.

We now give the same form and relate to other forms that have appeared in the literature.

Recall, if the areas are planned then area $i$ corresponds to Admin-1 strata $h_1[i]$, and if an unplanned (Admin-2) area then it is a subregion of $h_1[i]$.
The form given in \cite{sas2016sas} is,
$$ \widehat{V} (\widehat p_i^{\tiny{\,\mbox{HJ}}} ) = \sum_{h_2=1}^2 \widehat{V}(\widehat p_i^{\tiny{\,\mbox{HJ}}} ). $$
If $n_{h_1[i]h_s} > 1$, the variance is built from stratum centered residuals (and we take the  finite population correct factor as $f_{ih_1h_2}=0$, since we sample a small fraction of the available units in each strata),
\begin{equation}\label{eq:sasorig}
\widehat{V} (\widehat p_i^{\tiny{\,\mbox{HJ}}} ) = \frac{ n_{ih_1[i]h_2}}{n_{ih_1[i]h_2}-1} \sum_{c \in S_{ih_1[i]h_2} } (r_{ih_1[i]h_2c\cdot} - \overline{r}_{ih_1[i]h_2 \cdot \cdot})^2.
\end{equation}
 The residual components are 
\begin{eqnarray*}
r_{ih_1[i]h_2c \cdot}&=& \frac{1}{v_{ih_1[i]\cdot\cdot\cdot}}\sum_{k \in S_{ih_1[i]h_2c}}
v_{hck}(z_{hck} - \widehat p_i^{\tiny{\,\mbox{HJ}}} )\\
\overline{r}_{ih_1[i]h_2 \cdot \cdot} &=& \frac{1}{n_{ih_1[i]h_2}} \sum_{c \in S_{ih_1[i]h_2}} r_{ih_1[i]h_2c \cdot}
\end{eqnarray*}
Note that $n_{ih_1[i]h_2}$ is the number of samples in the planned domain and this is used even in the case of unplanned domains.

In \cite{lohr:10}, equation (6.34), the variance for a single strata and two-stage cluster sampling (relevant for a planned domain) is given as (with an obvious change in notation, i.e.,~dropping the strata indices):
\begin{equation}\label{eq:lohr}
\widehat V_{\text{Lohr}}(\widehat p_i^{\tiny{\mbox{HJ}}})
=
\frac{n_i}{n_i-1} 
\frac{1}{v_{i\cdot\cdot}^2}
\sum_{c \in S_i}
\left(
\sum_{k \in S_{ic}}
v_{ick}\bigl(z_{ick} - \widehat p_i^{\tiny{\mbox{HJ}}}\bigr)
\right)^2.
\end{equation}
Dropping the $h_1,h_2$ subscripts, (\ref{eq:sasorig}) is,
$$
\widehat{V} (\widehat p_i^{\tiny{\,\mbox{HJ}}} ) = \frac{n_i}{n_i-1} \sum_{c \in S_i } (r_{ic\cdot} - \overline{r}_{i\cdot \cdot})^2,
$$
where
\begin{eqnarray*}
r_{ic \cdot}&=& \frac{1}{v_{i\cdot\cdot}}\sum_{k \in S_{ic}}
v_{ick}(z_{ick} - \widehat p^{\tiny{\,\mbox{HJ}}} )\\
\overline{r}_{i \cdot \cdot} &=& \frac{1}{n_i} \sum_{c \in S_i} r_{ic \cdot}
\end{eqnarray*}
It is straightforward to show that in this case $\overline{r}_{i\cdot \cdot} =0$  and
so
$$ 
\widehat{V} (\widehat p_i^{\tiny{\,\mbox{HJ}}} ) = \frac{n_i}{n_i-1} \sum_{c \in S_i } r_{ic\cdot}^2  = \frac{n_i}{n_i-1} \frac{1}{v^2_{i\cdot\cdot}} \sum_{c \in S_i }\left(
\sum_{k \in S_{ic}}
v_{ick}(z_{ick} - \widehat p_i^{\tiny{\,\mbox{HJ}}} )\right)^2$$
which is identical to (\ref{eq:lohr}).

For a total in a generic area the estimator is,
$$
\widehat{T}_i = \sum_{h_2=1}^2\sum_{c \in S_{ih_1[i]h_2}} \sum_{k \in S_{ih_1[i]h_2c}} v_{ih_1[i]h_2ck} z_{ih_1[i]h_2ck}.
$$
The variance is \citep{sas2016sas}:
$$
\widehat V ( \widehat{T}_i ) = \sum_{h_2=1}^2
\widehat V_{h_2} ( \widehat{T}_i )
$$
where, again setting $f_{ih_1[i]h_2}=0$,
\begin{eqnarray*}
\widehat V_{h_2} ( \widehat{T}_i )
&=& 
\frac{n_{ih_1[i]h_2}}
{n_{ih_1[i]h_2}-1} 
\sum_{c \in S_{ih_1[i]h_2}}
(u_{ih_1[i]h_2c\cdot}-\overline{u}_{ih_1[i]h_2\cdot \cdot})^2\\
u_{ih_1[i]h_2c\cdot}&=& \sum_{k \in S_{ih_1[i]h_2c}} 
v_{ih_1[i]h_2ck} 
z_{ih_1[i]h_2ck}\\
\overline{u}_{h_1[i[h_2\cdot \cdot} &=& \frac{1}{n_{ih_1h_2}
} \sum_{c \in S_{ih_1[i]h_2}} u_{ih_1[i]h_2 c\cdot}
\end{eqnarray*}
When $H_1=H_2=1$, and simplifying notation accordingly:
$$
\widehat V ( \widehat{T}_i ) =\frac{n_i}{n_i-1}
\sum_{c \in S_i} \left[\left( \sum_{k\in S_{ic}} v_{ick}z_{ick}\right)-\frac{\widehat T_i}{n_i} \right]^2,
$$
which corresponds to equations (4) and (5) of \cite{lehtonen2009design}, for planned and unplanned domains, respectively.
Note that for unplanned domains, clusters in the larger planned domain (of which there are $n$) but not in the unplanned domain contribute $(\widehat T/n)_i^2 $ terms to the variance.
This form also corresponds to equation (4.6.2) in \cite{sarndal:etal:92}. All of these calculations are carried out under the assumption of sampling of PSUs (clusters) with replacement. This approximation is reasonable in the context of DHS surveys, since the fraction of clusters sampled from the master frame is small.

\section{Comparison of  Variance Estimators}

To recap, in the main paper we give the form of the variance that is appropriate for the DHS design, which uses a stratified, unequal probability two-stage cluster design.

Here, 
we compare estimates from this form
with those from the \texttt{survey} package using the 2018 Zambia DHS data. The {\tt survey} package uses variance formula from \cite{sarndal:etal:92}.

\begin{figure}[H]
\centering
\includegraphics[clip, trim=0.4cm 0cm 0.5cm 0cm, width=0.8\linewidth]{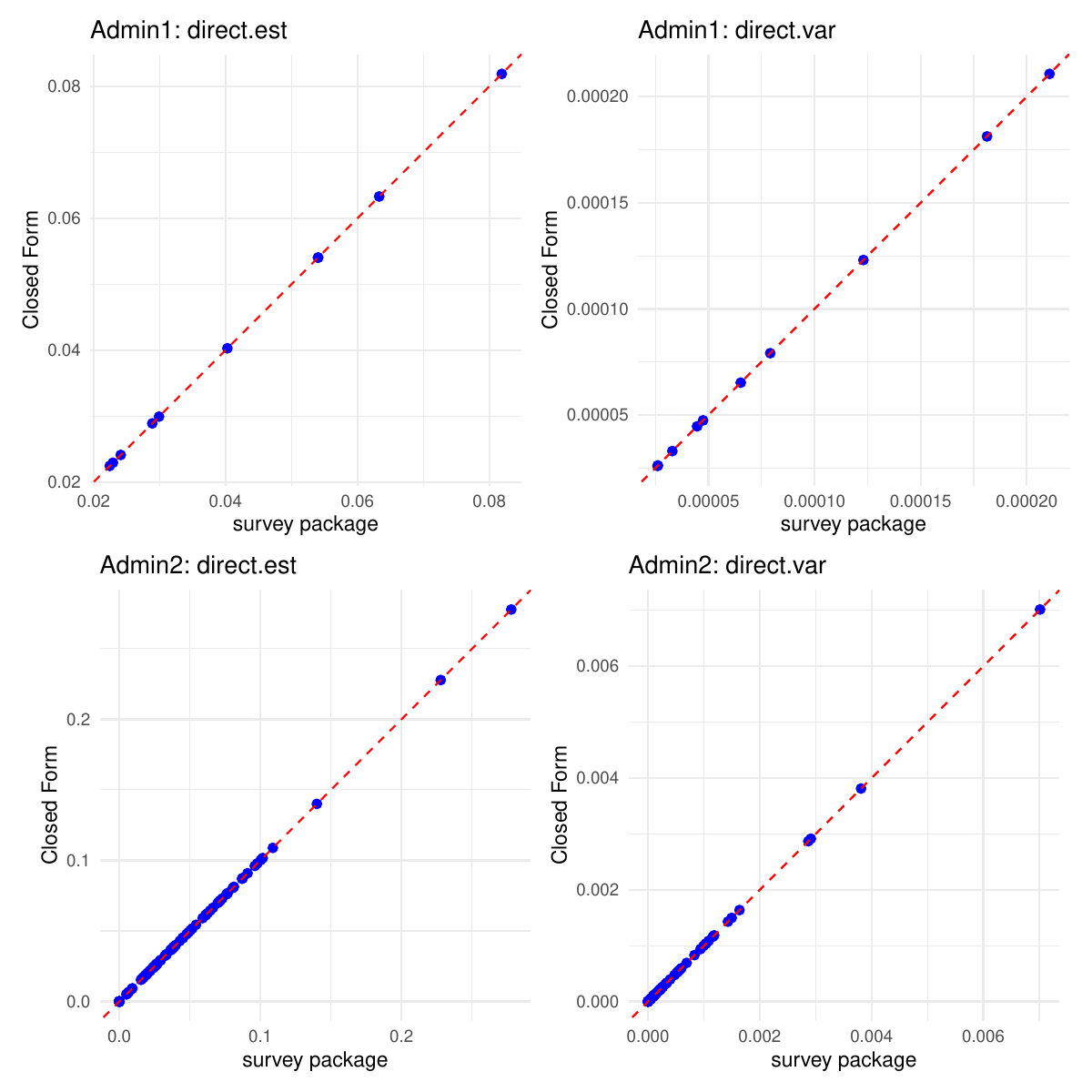}
\caption{%
Comparison of direct point estimates and variances computed using the \texttt{survey} package and closed-from formula for the wasting prevalence for children under five years of age (0-59 months) in Zambia, from the 2018 DHS.}
\label{fig:sas_survey_cmp}
\end{figure}

\vspace{0.5em}
Figure~\ref{fig:sas_survey_cmp} compares the direct estimates and corresponding variances 
from the two methods, at both Admin-1 and Admin-2 levels, and shows that the two results are almost identical, confirming the consistency of the two approaches.

\clearpage
\section{Additional Simulation Results}

\subsection{Large Sample Simulation Results}

We carry out another simulation study to assess the large sample properties of the proposed direct estimation
and variance-fixing procedures. Specifically, we conduct a large-sample simulation using the Admin-1 Central province in Zambia as our template.
The Central province consists of 10 Admin-2 districts, each stratified into urban and rural domains, though in this simulation we do not consider urban/rural areas. Hence, we have $H=10$ in this setting. 
Within each stratum, we randomly selected 50 clusters, and within each cluster we sampled 30 individuals following the DHS design. This yields a total of $(50 + 50) \times 10 = 1000$  clusters.

Individual-level binary responses were generated under a logistic mixed model
without Admin-2 random effects:
\[
\text{logit}(p_{ck}) = \text{logit}(m_0) + e_c + e_{ck}, \qquad
e_c \sim \mbox{N}(0, \sigma_1^2), \ e_{ck} \sim \mbox{N}(0, \sigma_2^2),
\]
where \(m_0 = 0.5\), \(\sigma_1 = 0.2\), and \(\sigma_2 = 0.05\).
Binary outcomes were then drawn as $$y_{ck} \mid p_{ck} \sim \text{Bernoulli}\left(\text{expit}(p_{ck})\right).$$ Here, no spatial or regional effects were introduced, ensuring that all variability arises
from the sampling design and within-cluster random variation.


The same variance-handling strategies were compared, to evaluate the estimation procedures:
\begin{enumerate}
    \item {\it All-non-fixed:} 
Use the variance estimate in \eqref{eq:sasvar} for all domains.
\item {\it All-Fixed:} 
Use variance adjustment formula  for all domains, 
regardless of whether the estimated variance is legal or not.
\item {\it Fixed (Illegal) and non-fixed (Legal):} 
Use variance estimate only for domains which have illegal variances, with \eqref{eq:sasvar} used for legal variances. 
\end{enumerate}

For each of the three metrics described in the paper, we compute its value within every simulation and then summarize performance by averaging over all iterations $s =1,\dots,S =1000$. We then compare the performance of the three strategies.
For each replication and each Admin-2 domain, we computed the performance metrics, as in Section 5 of the main paper.

\begin{figure}[H]
    \centering
    \includegraphics[width=\textwidth]{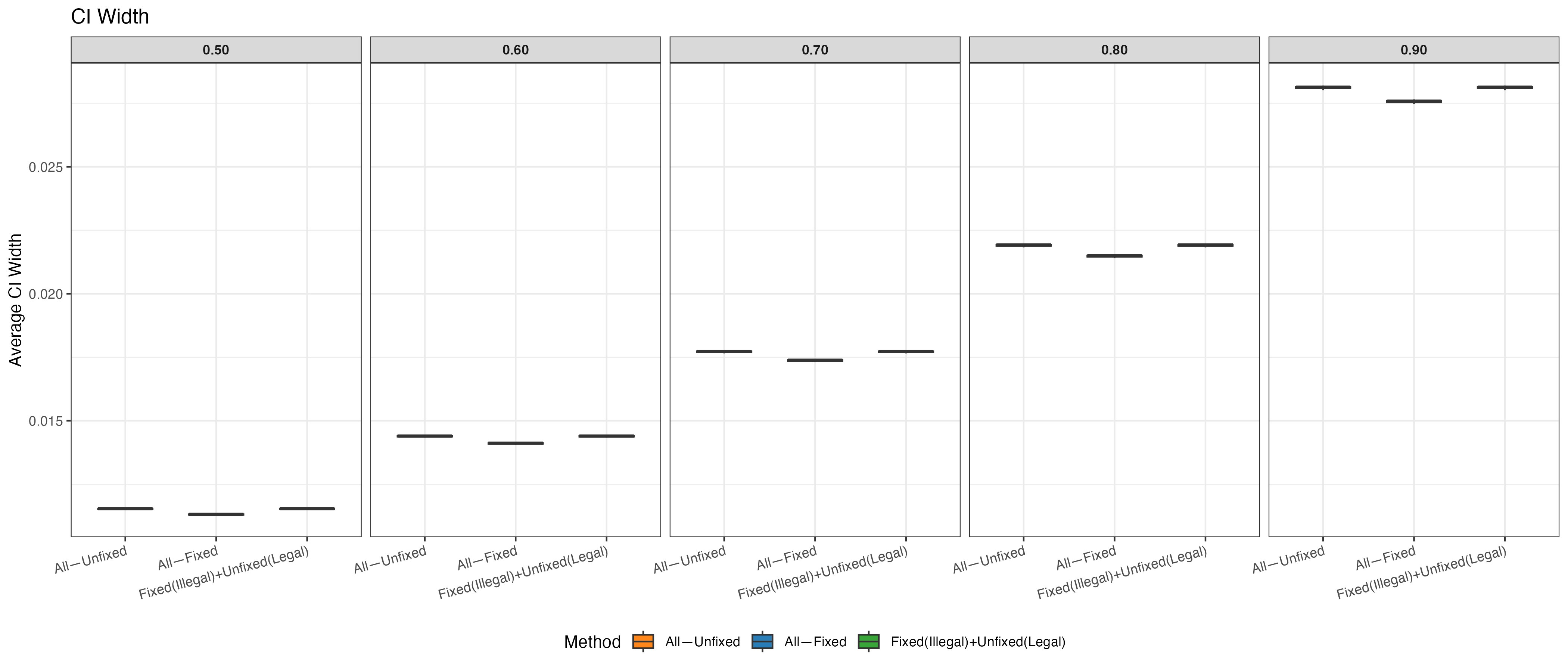}
    \caption{Average confidence interval width for different variance methods. Each column corresponds to a different nominal CI probability.}\label{fig:CIwidth}
\end{figure}

In Figure \ref{fig:CIwidth} we plot 100$(1-\alpha)\%$ CI widths, for different $\alpha$ values. The All Fixed method gives the narrowest intervals, as expected given the inclusion of prior phantom clusters increases information. The All non-fixed gives slightly wider intervals than the mixed method, again as expected.

\begin{figure}[H]
    \centering
    \includegraphics[width=\textwidth]{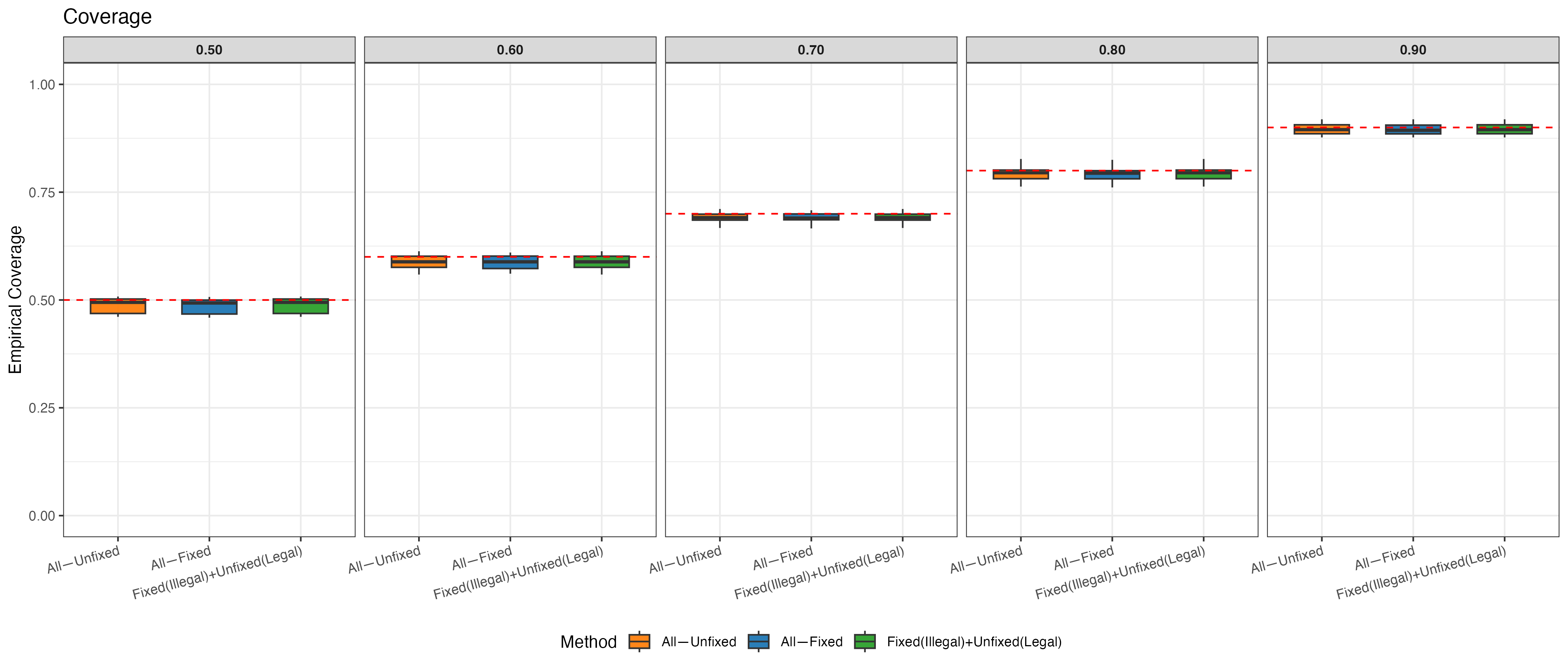}
       \caption{Empirical confidence interval coverage. Each column corresponds to a different nominal CI probability.}\label{fig:CIcov}
\end{figure}

In this large-sample setting, the variance estimates produced by all procedures closely track the true sampling variability, yielding empirical coverage rates that align well with their nominal levels (Figure \ref{fig:CIcov}). The variance-fixing adjustments have minimal impact here. 

\begin{figure}[H]
    \centering
    \includegraphics[width=\textwidth]{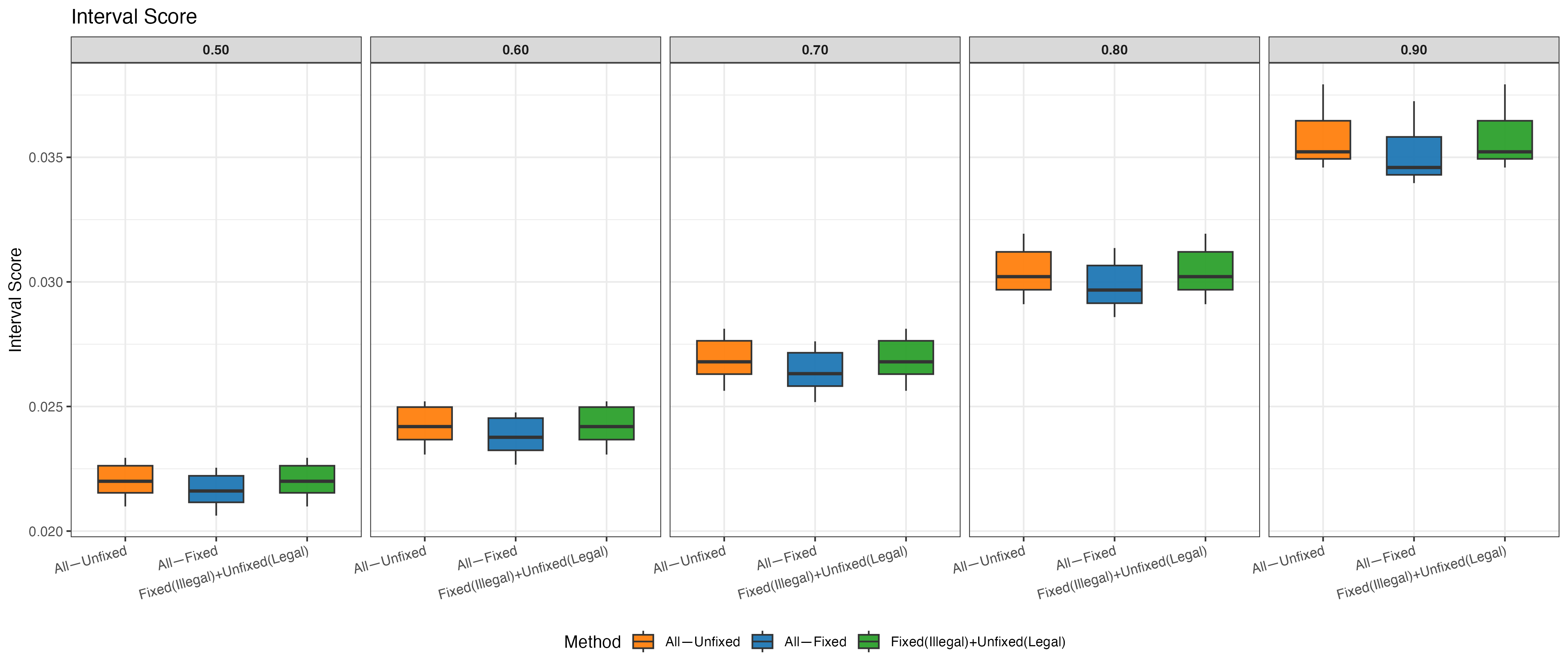}
          \caption{Interval scores for three methods. Each column corresponds to a different nominal CI probability.}\label{fig:IS}
\end{figure}

The interval score plot (Figure \ref{fig:IS} shows that the All Fixed method gives the most favorable scores, followed by the mixed method, which is slightly better than the All non-fixed approach. The differences are small though.

These results illustrate that our estimator remains stable and reliable when sample sizes are large. Impact from variance correction, while essential in small-sample scenarios, appropriately diminishes as the effective information increases.

\clearpage
\subsection{Further Simulation Results}

\begin{figure}[H]
    \centering
    \includegraphics[width=\textwidth]{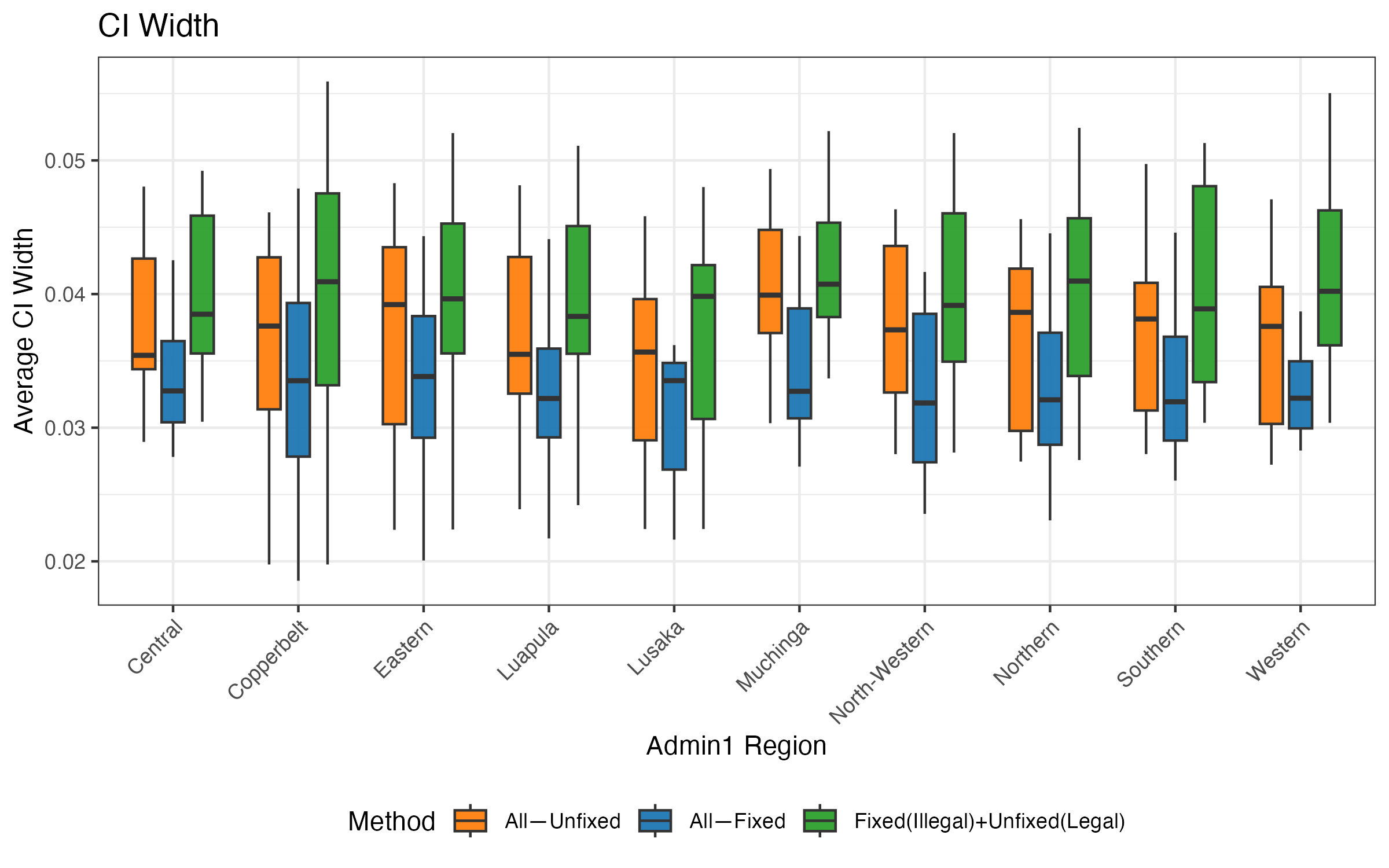}
    \caption{The averaged (across simulations) widths of 80\% confidence interval for Admin-2 areas, plotted as as a function of Admin-1 areas. 
    }
    \label{fig:ci_width_0.8}
\end{figure}

We present additional results for the simulation study detailed in Section 5 of the main paper.  In Figure \ref{fig:ci_width_0.8}, the CI widths are summarized and the results are as expected. The shortest intervals are when all areas are fixed (because of the effective increase in sample size from the phantom clusters), with the longest being when only illegal variances are fixed. The all non-fixed method includes the zero width intervals. 

\begin{figure}[H]
    \centering
    \includegraphics[width=0.95\textwidth]{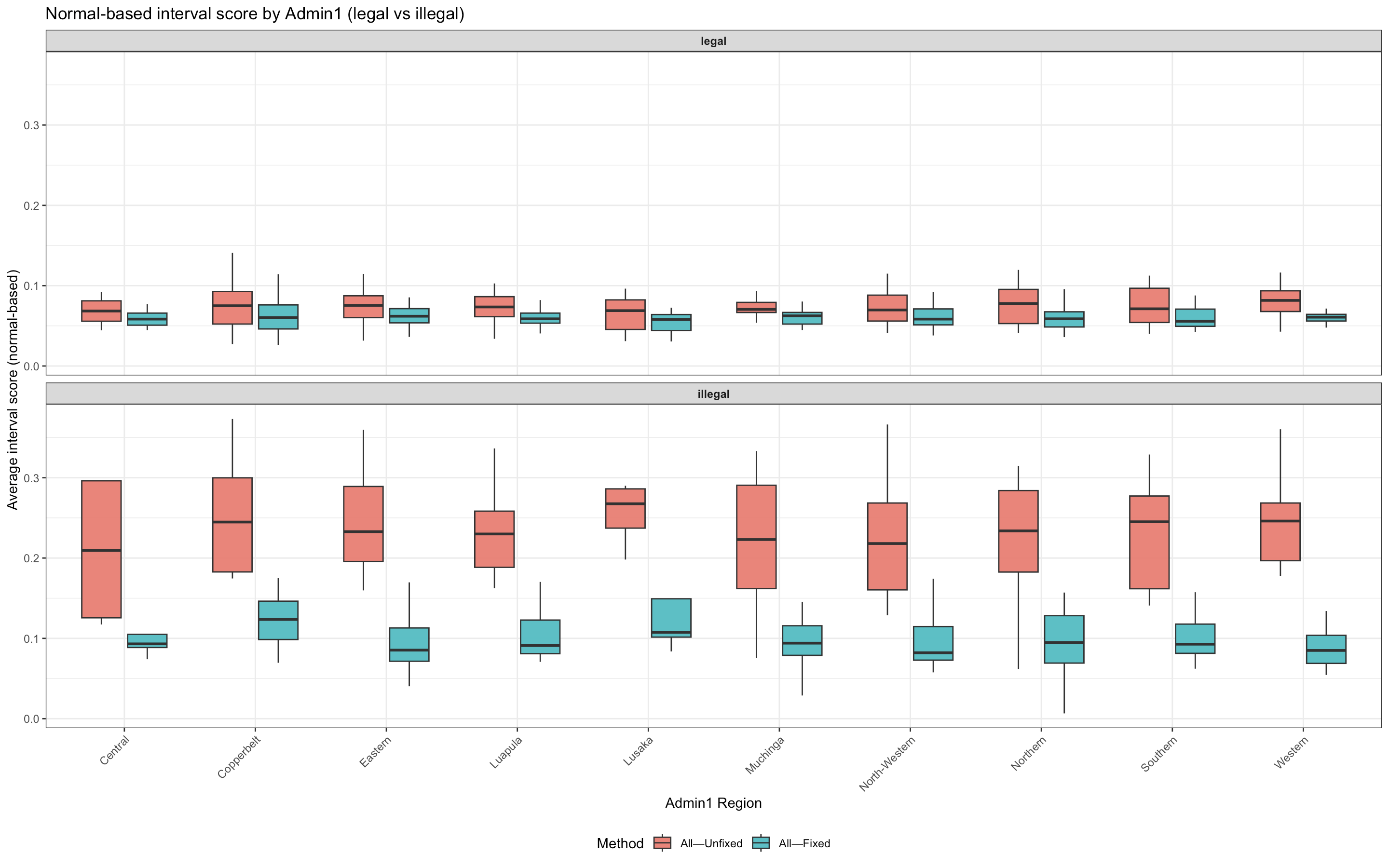}
    \caption{Interval scores (with 80\% coverage) for Admin-2 domains for fixed and non-fixed 
        variance-handling strategies, by Admin-1 domains and  for legal and illegal regions.
    }
    \label{fig:stratified_ISs}
\end{figure}

To further disentangle the sources of improvement resulting from variance adjustment, we compare interval scores under the fixed and non-fixed variance across regions classified as legal or illegal based on their variance estimates. As shown in Figure \ref{fig:stratified_ISs}, the primary gains arise within the illegal regions, where correcting the variance has the greatest impact.

\clearpage

\section{Further Results for Zambia}

\begin{figure}[H]
    \centering
        \includegraphics[clip, trim=0cm 5cm 0cm 4.5cm, width=1\linewidth]{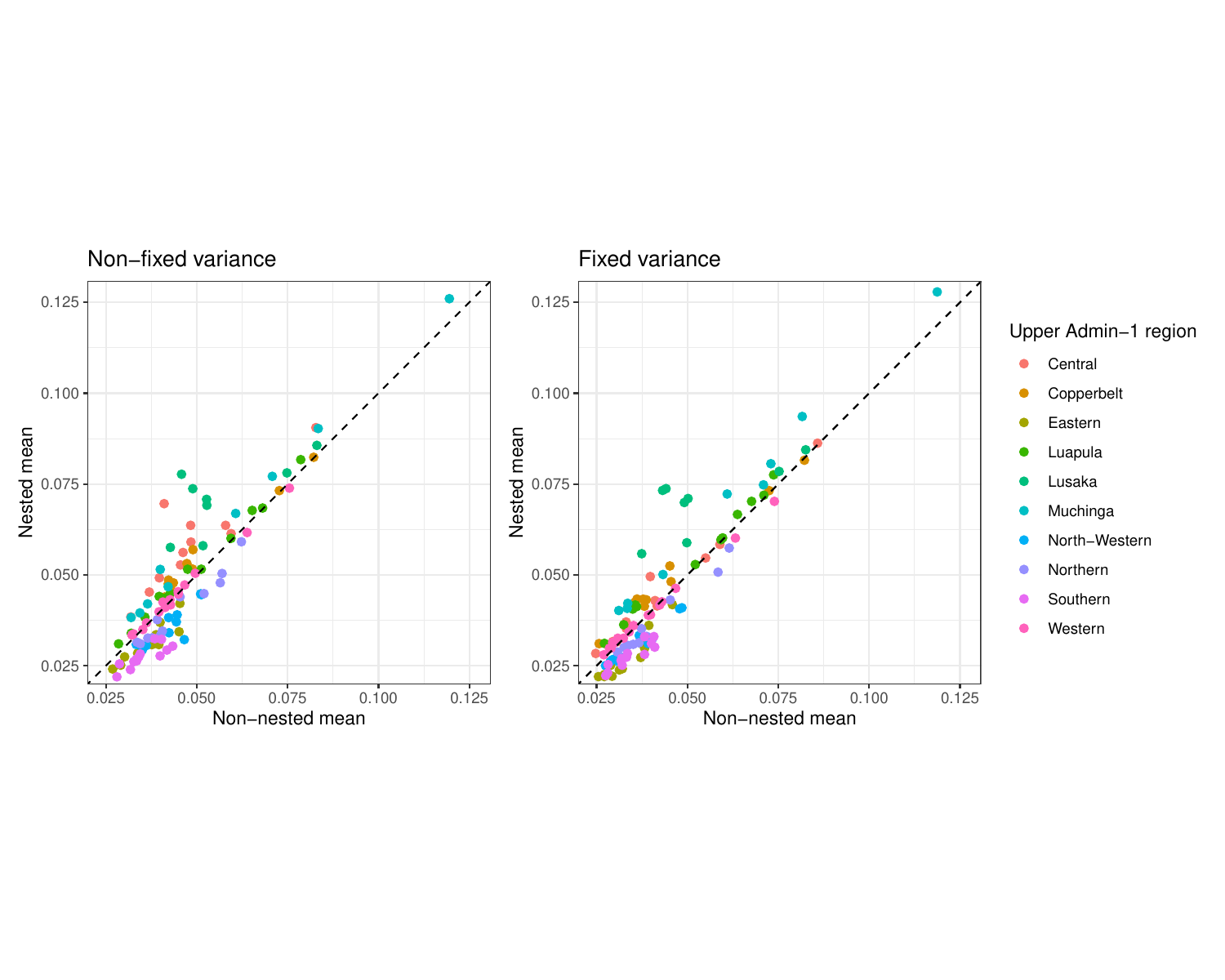}
        \caption{Prevalence estimates for Admin-2 areas in Zambia. In the left plot we compare non-nested and nested Fay-Herriot model estimates in the case when the problematic variance areas are not modified. In the right plot we consider the modified variance case.}
    \label{fig:nested_nonnested}
\end{figure}

In Figure \ref{fig:nested_nonnested} we compare Admin-2 prevalence point estimates from non-nested and nested Fay-Herriot models, for both the non-adjusted and adjusted models. The nested models include Admin-1 fixed effects, with BYM2 random effects modeling within Admin-1 variation. The spread of the estimates is greater under the nested models, which is desirable, since the aim of introducing the fixed effects is to reduce over-shrinkage.

\begin{figure}[H]
    \centering
            \includegraphics[clip, trim=0cm 2cm 0cm 3cm, width=1\linewidth]{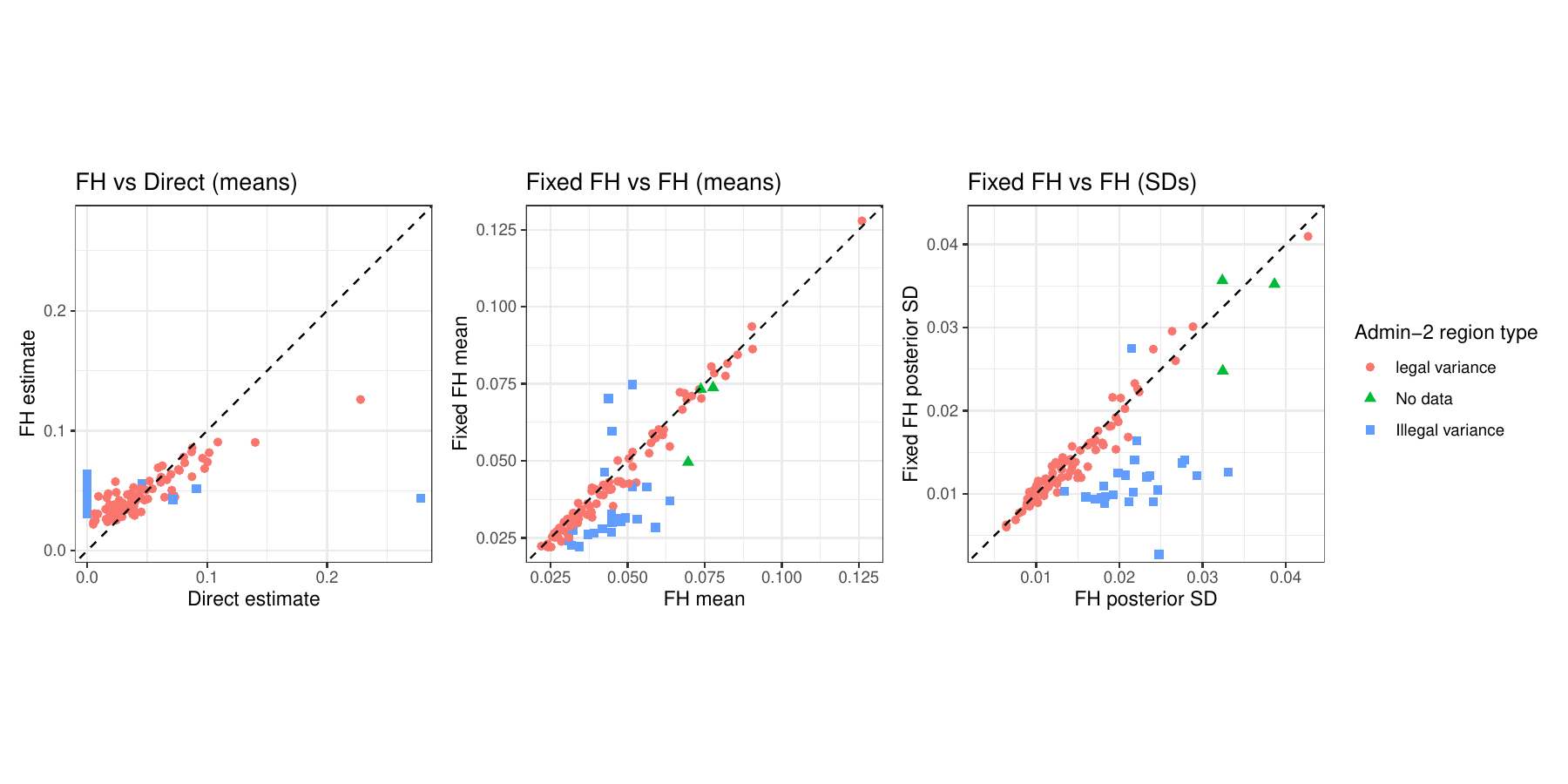}
    \caption{Comparison of direct estimates and Fay-Herriot estimates (two versions, with and without variance fix) for Admin-2 areas in Zambia. Left: direct vs Fay-Herriot estimates. Middle: Fay-Herriot means vs fixed Fay-Herriot means. Right: Fay-Herriot standard deviations vs fixed Fay-Herriot deviations.}
    \label{fig:scatter_compare0}
\end{figure}

In Figure \ref{fig:scatter_compare0} we compare the fixed and non-fixed Fay-Herriot estimates and standard deviations with each other, and with the direct estimates. The left hand panel shows the shrinkage of the Fay-Herriot estimates, relative to the direct estimates. Comparing the fixed Fay-Herriot estimates versus the non-fixed  Fay-Herriot estimates is interesting since we see the majority of the fixed estimates are smaller than the non-fixed. This is because the latter are treated as missing data and so are imputed from the model. The majority of the areas need fixing because there are no events (i.e.,~no children that are wasted). Hence, this is informative since it suggests that the prevalence in those areas is relatively low, which explains why the fixed versions are lower than the non-fixed. We also examine the uncertainty in the estimates and see that in general the uncertainty is lower in the fixed version, because there we are using the extra information that is ignored in the non-fixed version. The bias that is introduced by treating as missing data, is akin to that which would arise if values that fall below a certain level are dropped from an analysis.

In Table \ref{tab:hyper} we compare hyperparameter estimates from the models described above. 
The variance fixed models lead to a reduction in the proportion of the variation that is spatial parameter $\phi$. The variance fix is a form of smoothing, which reduces the spatial contribution.
The nested models models have slightly smaller overall residual variation than the non-nested models, because the fixed effects (Admin-1 intercepts) absorb some of the spatial variation.

\begin{table}[ht]
\centering
\begin{tabular}{l l r r r r r }
\hline
Method & Hyperparameter & Mean & SD & 2.5\% & Median & 97.5\%  \\
\hline
  \multirow{2}{*}{Nested} 
  & Precision for Admin-2 & 4.57 & 2.16 & 1.81 & 4.11 & 10.09  \\
  & Phi for Admin-2       & 0.54 & 0.26 & 0.06 & 0.56 & 0.95  \\
  \hline
\multirow{2}{*}{Non-nested} 
  & Precision for Admin-2 & 4.07 & 1.50 & 1.96 & 3.79 & 7.79  \\
  & Phi for Admin-2       & 0.45 & 0.25 & 0.04 & 0.44 & 0.90  \\ \hline
\multirow{2}{*}{Fixed Nested} 
  & Precision for Admin-2 & 5.08 & 1.89 & 2.46 & 4.74 & 9.77 \\
  & Phi for Admin-2       & 0.26 & 0.23 & 0.01 & 0.18 & 0.81  \\ \hline
\multirow{2}{*}{Fixed Non-nested} 
  & Precision for Admin-2 & 3.89 & 1.13 & 2.19 & 3.71 & 6.59  \\
  & Phi for Admin-2       & 0.23 & 0.21 & 0.01 & 0.17 & 0.74  \\
\hline
\end{tabular}
\caption{BYM2 hyperparameter summary for four spatial Fay-Herriot models. The precision is $\sigma_u^{-2}$, where $\sigma_u$ is the standard deviation parameter in the BYM2 model, and $\phi$ is the proportion of the variation that is spatial.}
\label{tab:hyper}
\end{table}

We now aggregate the Admin-2 estimates to Admin-1 and nataional levels, to further investigate properties of the models. 
For the nested models, we can obtain an Admin-1 estimate from the Admin-2 estimates as,
\begin{equation}\label{eq:wp}
p_{h_1} = \sum_{i=1}^{I_{h_1}} q_{h_1[i]} \times \mbox{expit}( \alpha_{h_1[i]} + u_i),
\end{equation}
where $I_{h_1}$ is the number of Admin-2 areas in Admin-1 area $h_1$ and $q_{h_1[i]i}$ is the fraction of Admin-2 area $i$ that lies in Admin-1 area $h_1[i]$. We obtain these fractions from WorldPop \citep{tatem2017worldpop}.
Note that we do not use a sum-to-zero constraint within the Admin-1 areas indexed by $h_1$.
We can approximate (\ref{eq:wp}) by:
$$
\widehat p_{h_1} = \sum_{i=1}^{I_{h_1}} q_{h_1[i]} \times \mbox{expit}( \widehat \alpha_{h_1[i]}+ \widehat u_i)
$$
where $\widehat \alpha_{h_1}$ and $\widehat u_i$ 
and the posterior medians. A more rigorous approach would be to use posterior samples for $\alpha_{h_1}$ and $u_i$.

The national prevalence is,
$$p = \sum_{h_1=1}^{10} q_{h_1} \times p_{h_1}, $$
where we may use WorldPop, or design weights to obtain the population fractions, with the latter given by,
\begin{eqnarray*}
q_{h_1} &=& \frac{\sum_{h_2=1}^2 \sum_{c \in S_{ih_1h_2ck}} \sum_{k \in S_{h_1h_2c}} w_{ih_1h_2ck}}{\sum_{h_1=1}^{10}\sum_{h_2=1}^2 \sum_{c \in S_{h_1h_2}} \sum_{k \in S_{h_1h_2c}} w_{ih_1h_2ck}}.
\end{eqnarray*}

In Table \ref{tab:placeholder}, we present national prevalence estimates under different methods. We take as gold standard the direct (weighted) estimate with aggregation fractions corresponding to the design weights. Using the WorldPop fractions raises the estimate by 0.0006, which is a relative percent increase of 1.4\%. The non-fixed nested estimates treat the areas with problematic variances as missing. But the data in these areas is informative (as discussed above), since it is suggestive that the prevalence is low, because for this application, zero events, giving prevalence estimates of zero, is a common issue in the missing areas. Hence, the aggregated nested estimate in the non-fixed case gives a relative increase of 10.4\% when aggregation is via survey weights and 12.3\% when via WorldPop fractions. The fixed models perform better with increases of 3.5\% and 5.0\% using weights and WorldPop, respectively.

\begin{table}
    \centering
    \begin{tabular}{l|cc|cc}
 & \multicolumn{2}{c|}{Survey Weight Fractions}&\multicolumn{2}{c}{WorldPop Fractions}\\ 
  Method &Estimate&95\% CI&Estimate &95\% CI \\ \hline
Aggregated, direct        & 0.0423 & (0.0370, 0.0486)&0.0429&(0.0373, 0.0493)\\
 Aggregated, non-fixed nested        &  0.0467 & (0.0415, 0.0525)&0.0475 &(0.0422, 0.0537)\\
  Aggregated, fixed nested       & 0.0438  & (0.0393, 0.0488)&0.0444&(0.0399, 0.0498)\\
    \end{tabular}
    \caption{National prevalence estimates under different methods.}
    \label{tab:placeholder}
\end{table}

\noindent 

\begin{figure}[H]
    \centering
            \includegraphics[clip, trim=0cm 2cm 0cm 3cm, width=1\linewidth]{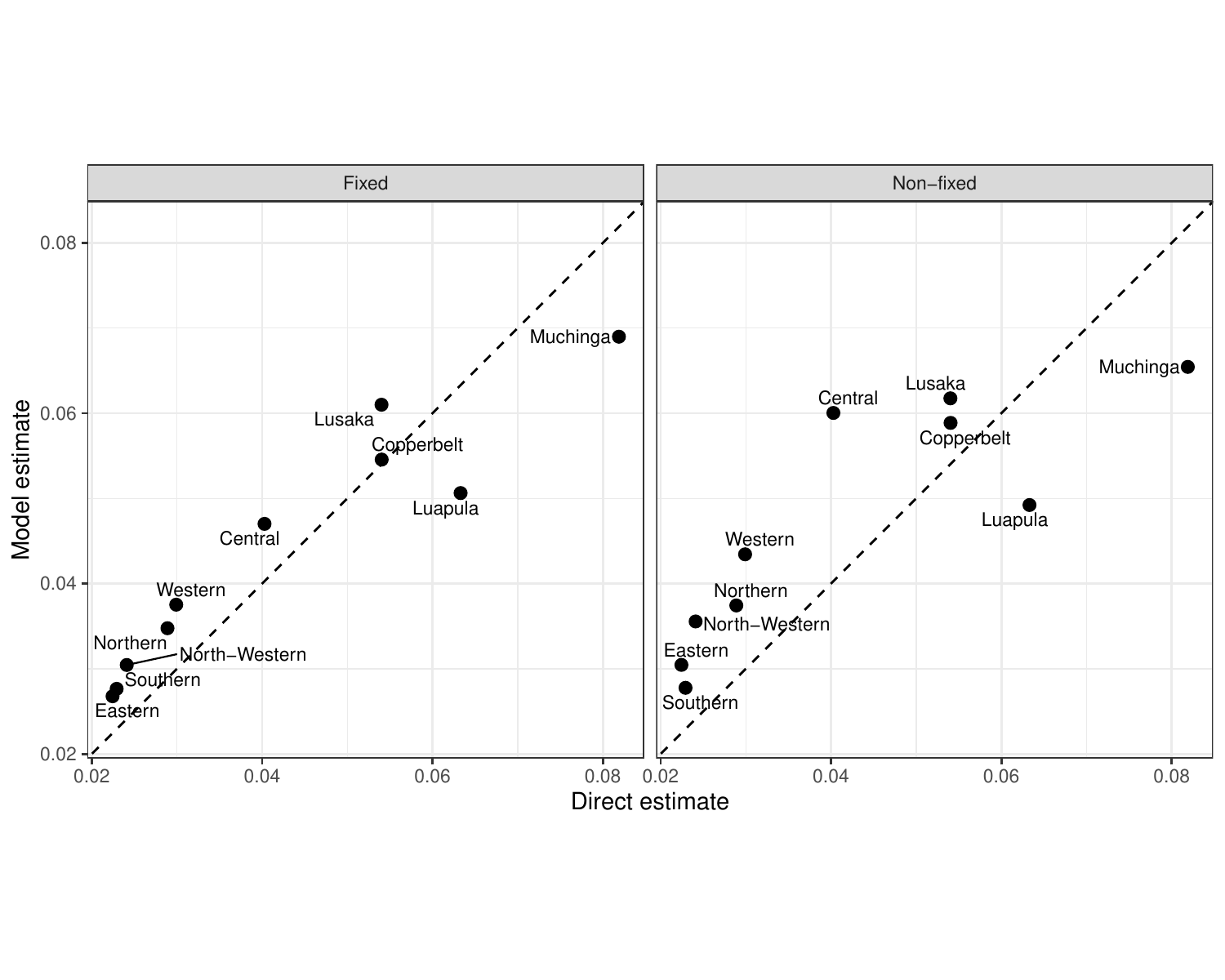}
    \caption{Fixed and non-fixed Fay-Herriot posterior mean estimates (using nested models) plotted against direct estimates.}
    \label{fig:scatter_compare1}
\end{figure}

Figure \ref{fig:scatter_compare1} shows that the Admin-1 aggregated estimates under a fixed Admin-2 Fay-Herriot model are closer to the direct estimates than are the non-fixed estimates. The non-fixed estimates are generally too large, in line with the previous discussion. Table \ref{tab:adm1_prev} gives the numerical values.

\begin{table}[ht]
\centering
\begin{tabular}{lccc}
\hline
Region & Direct estimate & Nested Non-Fixed & Nested Fixed \\
\hline
Central        & 0.0403 (0.0260, 0.0618) & 0.0597 (0.0418, 0.0835) & 0.0466 (0.0350, 0.0609) \\
Copperbelt     & 0.0540 (0.0402, 0.0722) & 0.0586 (0.0447, 0.0754) & 0.0545 (0.0424, 0.0693) \\
Eastern        & 0.0224 (0.0144, 0.0349) & 0.0300 (0.0204, 0.0418) & 0.0265 (0.0194, 0.0350) \\
Luapula        & 0.0633 (0.0415, 0.0954) & 0.0494 (0.0381, 0.0634) & 0.0510 (0.0401, 0.0631) \\
Lusaka         & 0.0540 (0.0360, 0.0804) & 0.0613 (0.0434, 0.0851) & 0.0607 (0.0419, 0.0874) \\
Muchinga       & 0.0819 (0.0576, 0.1152) & 0.0655 (0.0479, 0.0890) & 0.0685 (0.0524, 0.0903) \\
Northern       & 0.0289 (0.0204, 0.0408) & 0.0376 (0.0285, 0.0492) & 0.0346 (0.0270, 0.0438) \\
North Western  & 0.0241 (0.0140, 0.0413) & 0.0358 (0.0212, 0.0549) & 0.0302 (0.0209, 0.0431) \\
Southern       & 0.0229 (0.0140, 0.0373) & 0.0279 (0.0196, 0.0383) & 0.0280 (0.0189, 0.0403) \\
Western        & 0.0299 (0.0190, 0.0468) & 0.0436 (0.0308, 0.0585) & 0.0373 (0.0286, 0.0483) \\
\hline
\end{tabular}
\caption{Admin-1 estimates under direct and non-fixed and fixed nested Fay--Herriot models.}
\label{tab:adm1_prev}
\end{table}

\clearpage

\end{document}